\documentclass[reprint,amsmath,amssymb,amsbsy,superscriptaddress,pra,footinbib]{revtex4-1}
\usepackage{graphicx}
\usepackage{braket}
\usepackage{bm}
\usepackage{color}
\usepackage{ulem}
\usepackage{verbatim}
\usepackage[colorlinks=true, citecolor=blue, linkcolor=blue]{hyperref}
\usepackage{times}

{\par\noindent{\em #1:\ }}%
{~\rule{2mm}{2mm}\par\bigskip}

\begin{document}
\title{Bulk-edge correspondence in two-dimensional topological semimetals: 
A transfer matrix study of antichiral edge modes} 

\author{Tomonari Mizoguchi}
\affiliation{Department of Physics, University of Tsukuba, Tsukuba, Ibaraki 305-8571, Japan} 
\email{mizoguchi@rhodia.ph.tsukuba.ac.jp}
\author{Tohru Koma}
\affiliation{Department of Physics, Gakushuin University, Mejiro, Toshima-ku, Tokyo 171-8588, Japan}
\email{tohru.koma@gakushuin.ac.jp}

\date{\today}
\begin{abstract}
We study edge modes in topological semimetals which have an energy band structure of ordinary
semimetals but can be characterized by a Chern number. 
More specifically, we focus on a Qi-Wu-Zhang-type square-lattice model and a Haldane-type honeycomb model,
both of which exhibit antichiral edge modes whose wave packets propagate in the same direction 
at both parallel edges of the strip. 
To obtain these analytical solutions of the edge modes, 
we apply the transfer matrix method which was developed in the previous work 
[Phys. Rev. B \textbf{101}, 014442 (2020)]. 
As a result, we show that the bulk-edge correspondence is broken down 
for a certain range of the model parameters. 
More precisely, when increasing the strength of
a hopping amplitude of the Qi-Wu-Zhang-type model, the edge modes abruptly disappear, although the non-trivial
Chern number does not change.
In the Haldane-type model, for varying the model parameters, the edge modes do not 
necessarily disappear, and the non-trivial Chern number does not change.  
However, the energy spectral flows of the edge modes from the valence band to the conduction band 
are abruptly broken at a certain set of the model parameters.
\end{abstract}

\maketitle
\section{Introduction} 
As is well known, in certain band insulators, gapless modes can be found to be localized 
at a surface of the sample, although the bulk energy gap exists above the valence band.  
Unfortunately, those surface states are often unstable against perturbations, such as disorders 
of the surface, which are inevitable in experiments. 
In contrast to those ordinary band insulators, topological insulators (TIs)~\cite{Hasan2010,Qi2011} are expected to show 
the remarkable robustness of the boundary modes against perturbations. 
This robustness is a consequence of the topological nature of the insulators 
which is characterized by a topological invariant defined for the gapped ground state of the bulk. 
The relation between the topological invariant of the bulk and the boundary modes is known as 
the bulk-boundary correspondence~\cite{Halperin1982,Hatsugai1993}. 
This also asserts that a nontrivial topological invariant implies the existence of a nontrivial boundary mode. 
However, the bulk-boundary correspondence for 
disordered systems was mathematically justified only in a few cases in one and two dimensions. 
(See, e.g., Refs.~\onlinecite{Elbau2002, Mizoguchi2019}.)

It has been recognized that the hosts of 
the boundary states are not necessarily insulating.
Actually, semimetallic systems can also possess boundary states~\cite{Hsieh2008,Hsieh2009,Mao2011,Xu2014,Ying2018,Bahari2019}.
Additionally, boundary states under semimetallic band structures are also found in a Floquet system~\cite{Upreti2019}.
On the other hand, Dirac and Weyl semimetals~\cite{Murakami2007,Vafek2014,Armitage2018} 
and nodal line semimetals~\cite{Fang2015,Yamakage2016} have attracted 
considerable interests recently. In these systems, the conduction and valence bands touch 
at some points or lines. Clearly, the energy dispersions of these systems are totally different from 
those of ordinary semimetals. To definitely distinguish these two types of the energy dispersions, 
we recall the following: In both band insulators and ordinary semimetals, the conduction and valence 
bands are separated on the momentum space by a region where no electron states can exist. 
An energy gap between the two bands can be found in band insulators, while a range of energies 
in the conduction band overlaps with that in the valence band in ordinary semimetals. 
In the present paper, we will focus on certain topological semimetals that have the same type of 
energy dispersion as those of the above ordinary semimetals but they can be characterized by a Chern number.  
In the following, we will refer to generic semimetals that have the band structure of
the above-mentioned ordinary semimetals as semimetals for short.

 Among the phenomena occurring in semimetals, the emergence of antichiral edge modes 
is one of the most interesting phenomena~\cite{Colomes2018}. 
The wave packets of the modes propagate in the same direction at both parallel edges of the strip geometry in two dimensions. 
Their realizations have been 
investigated in many solid-state materials, e.g., 
transition metal dichalcogenides~\cite{Colomes2018,Vila2019}, 
exciton-polariton systems~\cite{Mandal2019},
a graphene-superconductor junction~\cite{Wang2020}, magnetic systems~\cite{Bhowmick2020}, 
and twisted bilayer graphene~\cite{Denner2020}. 
Furthermore, the antichiral edge modes are  
indeed observed in artificial materials such as 
electric circuits~\cite{Yang2020} and the gyromagnetic photonic crystals~\cite{Chen2020,Zhou2020}.

Despite these theoretical and experimental developments, 
the bulk-boundary correspondence in semimetals has not been understood sufficiently.
This is in sharp contrast to TIs, where the bulk-boundary correspondence has been well-established through many examples, 
and rigorous proof is given for some Altland-Zirnbauer classes~\cite{Hatsugai1993,Prodan2016,Graf2018,Mizoguchi2019}.
If the emergence of edge modes in semimetals is a consequence of the topological nature similarly to that of TIs, 
then one can expect the robustness of the edge modes against disorder and interactions. 
More specifically, the following question arises:
Does a non-trivial topological number of valence bands in semimetals guarantee the existence of edge modes?
In the following, we refer to semimetals that have a non-trivial topological number as topological semimetals (TSMs).

In this paper, we address this issue by 
using the transfer matrix method~\cite{Lee1981,Hatsugai1993_2,Molinari1997,Schulz-Baldes2000,Kellendonk2002,Elbau2002,Teo2008,Mao2010,Mao2011,Huang2012,Doh2014,Dwivedi2016,Pantaleon2017,Kunst2019,Mizoguchi2020}.
In general, the transfer matrix method is applicable regardless of the phase of the bulk, 
and thus is suitable for studying the bulk-boundary correspondence.
Yet, the exact solution of the boundary modes is accessible in only a few models.
In this regard, in Ref.~\onlinecite{Mizoguchi2020},
we developed the method to obtain the exact edge solutions
of arbitrary tight-binding models whose transfer matrix has the form of a $4 \times 4$ matrix.
By using this method, we study two concrete examples of the TSMs in two dimensions, namely, 
the Qi-Wu-Zhang (QWZ)-type model~\cite{Qi2006} and the Haldane-type model~\cite{Haldane1988}.
We find that the existence of the edge modes is not necessarily guaranteed by the non-trivial topological number,
but it depends on the parameters of the model.
This indicates that the bulk-edge correspondence does not hold for the TSMs.
 
The rest of this paper is organized as follows.
The main results of this paper are presented
in Sec.~\ref{sec:GQWZ} and Sec.~\ref{sec:GH},
where we investigate the QWZ-type model and the Haldane-type model, respectively.
In these two sections, we first discuss the bulk properties, and then we investigate the existence of the edge mode.
In Sec.~\ref{sec:summary}, we present a summary of this paper.
In Appendix.~\ref{sec:TM_method}, we review our method for obtaining exact solutions of edge modes 
on the basis of the transfer matrix method, which we have developed in Ref.~\onlinecite{Mizoguchi2020}.
In Appendix~\ref{sec:disp_exact}, we show
the concrete expressions of the exact solutions of the edge modes for the QWZ-type model and the Haldane-type model.

\section{Result 1: Qi-Wu-Zhang-type model \label{sec:GQWZ}}
We first study the QWZ-type model~\cite{Qi2006}.
The Hamiltonian is defined on a square lattice with $L_x \times L_y$ sites, 
and the fermions considered here have spin degrees of freedom, $\sigma = 1,2$. 
The Hamiltonian on a cylinder, where the open (periodic) boundary condition is imposed in the $x$ ($y$) direction,
reads
\begin{eqnarray}
H_{\rm QWZ} = H_{0} + H_{\rm SOI}, \label{eq:Ham_QWZ}
\end{eqnarray}
where 
\begin{subequations}
\begin{eqnarray}
H_{0}  =& t_1\sum_{\ell = 1}^{L_x-1} \sum_{m = 1}^{L_y} \sum_{\sigma} 
c^{\dagger}_{(\ell,m),\sigma} c_{(\ell + 1,m),\sigma}  + (\mathrm{h.c.}) \nonumber \\
-i &t_2 \sum_{\ell = 1}^{L_x} \sum_{m = 1}^{L_y} \sum_{\sigma} 
c^{\dagger}_{(\ell,m),\sigma} c_{(\ell ,m  +1),\sigma}  + (\mathrm{h.c.}), \nonumber \\
\end{eqnarray}
and 
\begin{widetext}
\begin{eqnarray}
H_{\rm SOI} &=&-i \alpha_1 \sum_{\ell = 1}^{L_x-1} 
\sum_{m = 1}^{L_y} \sum_{\sigma,\sigma^\prime} [\tau_1]_{\sigma,\sigma^\prime} 
c^{\dagger}_{(\ell, m), \sigma} c_{(\ell + 1, m), \sigma^\prime} + (\mathrm{h.c.}) \nonumber \\
&-&i \alpha_2   \sum_{\ell = 1}^{L_x} 
\sum_{m = 1}^{L_y} \sum_{\sigma,\sigma^\prime} [\tau_2]_{\sigma,\sigma^\prime} 
c^{\dagger}_{(\ell, m), \sigma} c_{(\ell, m + 1), \sigma^\prime} + (\mathrm{h.c.})\nonumber \\
&+&\alpha_3  \sum_{\ell = 1}^{L_x-1} 
\sum_{m = 1}^{L_y} \sum_{\sigma,\sigma^\prime} [\tau_3]_{\sigma,\sigma^\prime} 
c^{\dagger}_{(\ell, m), \sigma} c_{(\ell + 1, m ), \sigma^\prime} + (\mathrm{h.c.}) \nonumber \\
&+&\alpha_3  \sum_{\ell = 1}^{L_x} 
\sum_{m = 1}^{L_y} \sum_{\sigma,\sigma^\prime} [\tau_3]_{\sigma,\sigma^\prime} 
c^{\dagger}_{(\ell, m), \sigma} c_{(\ell, m + 1), \sigma^\prime} + (\mathrm{h.c.}) \nonumber \\
 &+&\sum_{\ell = 1}^{L_x} 
\sum_{m = 1}^{L_y} \sum_{\sigma,\sigma^\prime}
m_0 \alpha_3 [\tau_3]_{\sigma,\sigma^\prime} c^\dagger_{(\ell,m),\sigma} c_{(\ell,m),\sigma^\prime}.
\end{eqnarray}
\end{widetext}
\end{subequations}
Here $\ell$ and $m$ are the coordinates of the sites in the $x$ and $y$ directions, respectively,
and $c_{(\ell ,m),\sigma}$ denotes the annihilation operator of the fermion at the site $(\ell ,m)$ with spin $\sigma$.
The parameters $t_1$, $t_2$, $\alpha_\rho$ ($\rho=1,2,3$) and $m_0$ are real, and 
$\tau_\rho$ ($\rho=1,2,3$) stands for Pauli matrices.
Note that only the term $H_{\rm SOI}$ is often called the QWZ model. 
In fact, the topological semimetal can be realized due to $H_{0}$, as we will show below.  
\begin{figure}[t]
\begin{center}
\includegraphics[clip,width = \linewidth]{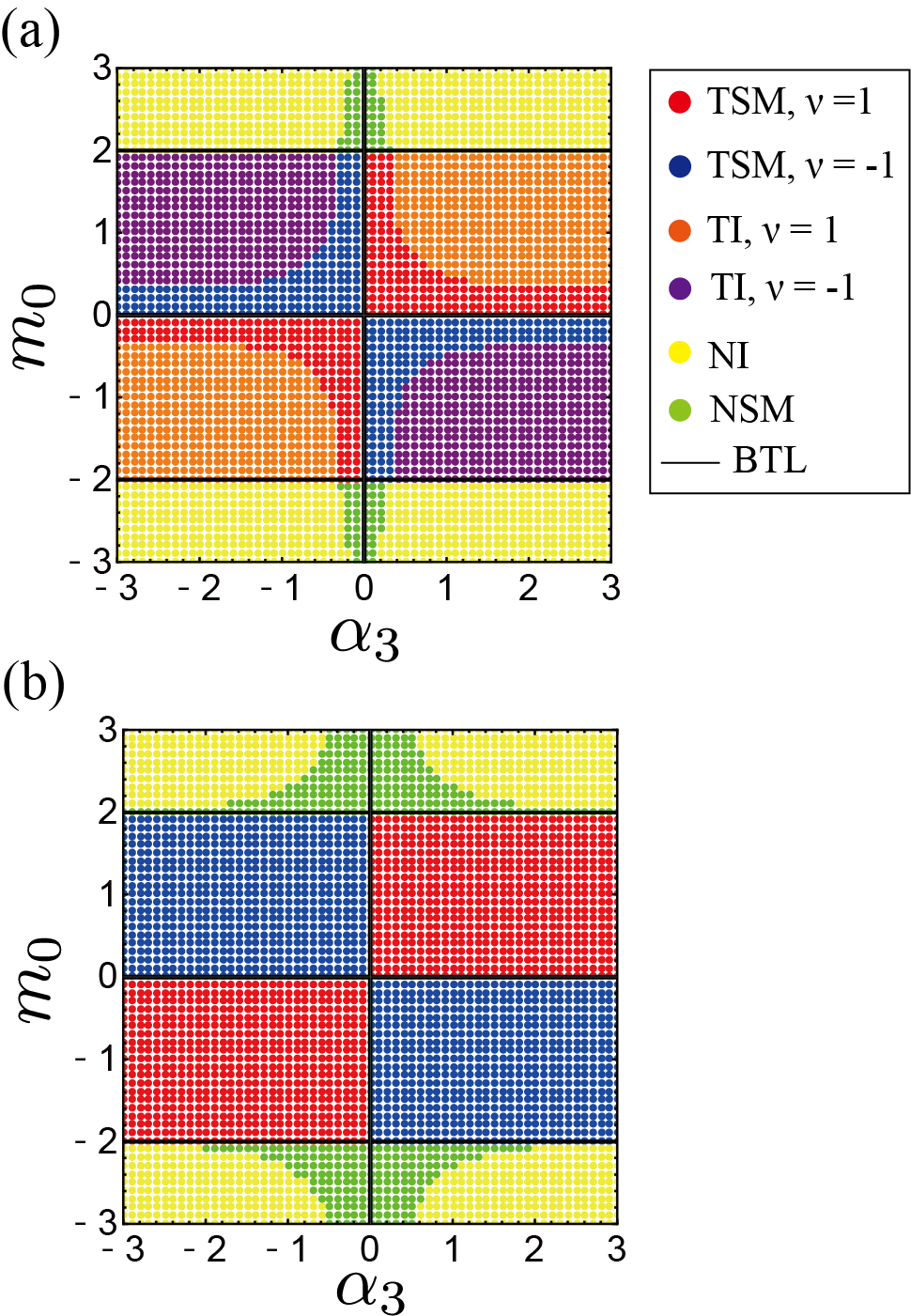}
\caption{The phase diagram of the QWZ-type model 
for (a) $t_1=-0.3$, $t_2=-0.3$, $\alpha_1 = \alpha_2 = 0.5$
and (b) $t_1=-0.5$, $t_2=-1$, $\alpha_1 = \alpha_2 = 0.5$.
Black lines denote the BTL.}
\label{fig:pd_qwz}
\end{center}
\end{figure}

\begin{figure*}[t]
\begin{center}
\includegraphics[clip,width = 0.98\linewidth]{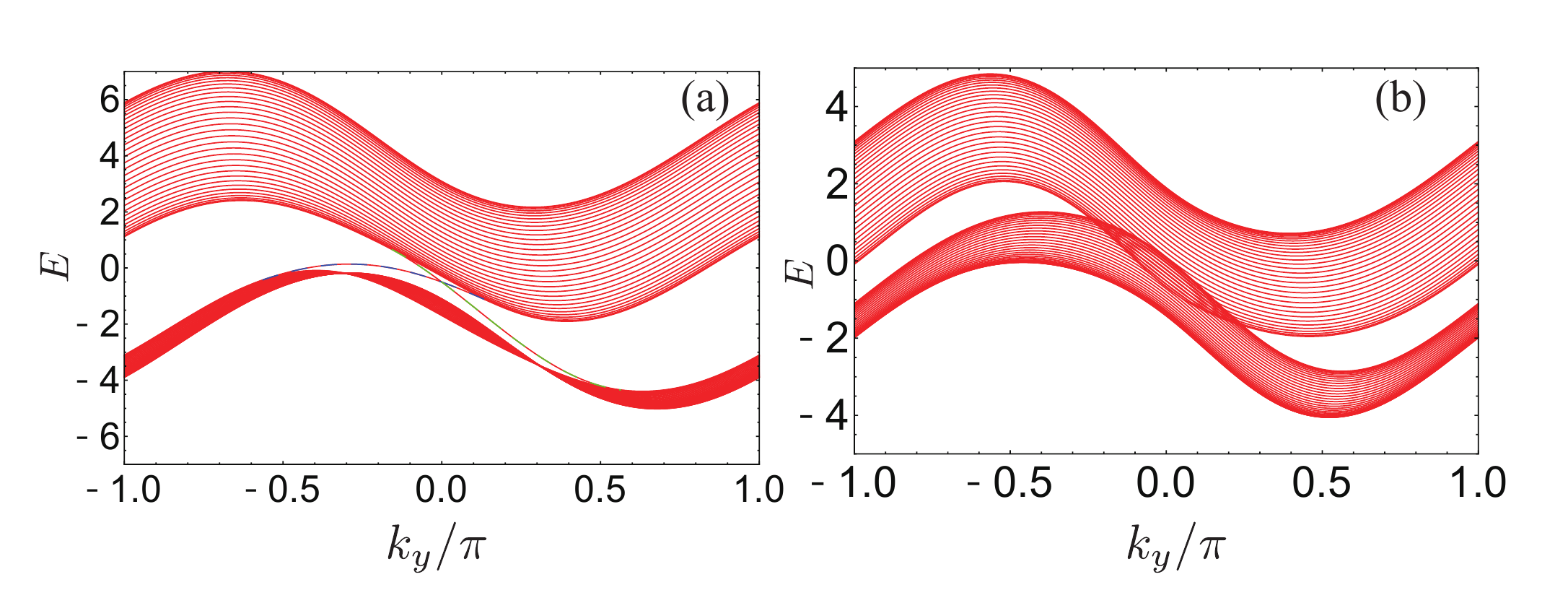}
\caption{The band structures for the QWZ-type model
on a cylinder
with 
$(t_1,t_2,\alpha_1,\alpha_2,\alpha_3,m_0) = $ 
(a) $(-0.5,-1,0.5,0.5,0.7,-1.5)$
and (b) $(-0.5,-1,0.5,0.5,0.3,-1.5)$.
Note that both of these panels are for the TSM phase. 
Red solid lines are obtained by the numerical diagonalization for $L_x=32$,
and blue and green broken lines are the exact solutions for the left and right edge modes, respectively.}
\label{fig:band_qwz}
\end{center}
\end{figure*}
\subsection{Bulk properties}
Before proceeding to the analysis of the edge modes,
let us summarize the bulk properties of the present model. 
We impose the periodic boundary condition in both the $x$ and $y$ directions,
and we have the Fourier transform,
\begin{eqnarray}
c_{(k_x,k_y),\sigma} = \frac{1}{\sqrt{L_x L_y}} \sum_{\ell=1}^{L_x} \sum_{\ell=1}^{L_y} 
e^{-i \left( k_x \ell  + k_y m \right) } c_{(\ell,m),\sigma}.
\end{eqnarray}
Then, the bulk Hamiltonian can be written as 
\begin{eqnarray}
H_{\rm QWZ} = \sum_{k_x,k_y} 
\bm{\Psi}^\dagger(k_x,k_y)
\mathcal{H}_{\rm QWZ}(k_x,k_y) 
\bm{\Psi}(k_x,k_y), \nonumber \\
\end{eqnarray}
where $\bm{\Psi}(k_x,k_y) = \left(c_{(k_x,k_y),1}, c_{(k_x,k_y),2}\right)^{\rm T}$ and 
\begin{eqnarray}
\mathcal{H}_{\rm QWZ}(k_x,k_y) = R_0 (k_x,k_y) I_2 + \bm{R}(k_x,k_y) \cdot \bm{\tau}. \label{eq:ham_QWZ_mom}
\end{eqnarray}
Here we have introduced
\begin{subequations}
\begin{eqnarray}
R_0 (k_x,k_y) = 2t_1 \cos k_x + 2t_2 \sin k_y,
\end{eqnarray}
\begin{eqnarray}
R_1 (k_x,k_y) = 2 \alpha_1 \sin k_x, \label{eq:Rx_qwz}
\end{eqnarray}
\begin{eqnarray}
R_2 (k_x,k_y) = 2 \alpha_2 \sin k_y, \label{eq:Ry_qwz}
\end{eqnarray}
and 
\begin{eqnarray}
R_3 (k_x ,k_y) = 2 \alpha_3 (m_0 + \cos k_x + \cos k_y).  \label{eq:Rz_qwz}
\end{eqnarray}
\end{subequations}
The energy eigenvalues of the two bands are given by 
\begin{eqnarray}
E_{\pm} (k_x,k_y)= R_0(k_x,k_y) \pm |\bm{R}(k_x,k_y)|. \label{eq:en_QW}
\end{eqnarray}
Clearly, when $|{\bf R}(k_x,k_y)|$ is nonvanishing, the two bands are separated by 
the forbidden region as mentioned in the
Introduction. Let us consider the condition that 
the two bands touch at some $(k_x, k_y)$ in the Brillouin zone. 
For simplicity, we assume $\alpha_\rho \neq 0$ for all $\rho=1,2,3$. 
Clearly, from Eq.~(\ref{eq:en_QW}), 
the condition implies $|{\bf R}(k_x,k_y)|=0$ for some $(k_x, k_y)$. 
From Eqs.~(\ref{eq:Rx_qwz}) and (\ref{eq:Ry_qwz}), this occurs only at $(k_x, k_y)=(0,0)$, $(\pi, 0)$, $(0,\pi)$, or $(\pi, \pi)$.  
Further, from $R_3(k_x, k_y)=0$, one has $m_0=-2$ for $(k_x, k_y)=(0,0)$, 
$m_0=0$ for $(k_x, k_y)=(\pi, 0)$ or $(0,\pi)$, and $m_0=2$ for $(k_x, k_y)=(\pi, \pi)$. 

If the band touching does not occur, the topological Chern number 
for the valence band is well-defined, and it is given as~\cite{Yakovenko1990,Qi2006}
\begin{eqnarray}
\nu = \frac{1}{4\pi} \int_{-\pi}^\pi \hspace{0.5mm} dk_x \int_{-\pi}^\pi \hspace{0.5mm} dk_y \hspace{0.5mm} \hat{R}(\bm{k}) \cdot 
\left[ \frac{\partial \hat{R}(\bm{k}) }{\partial k_x} \times \frac{\partial \hat{R}(\bm{k}) }{\partial k_y}\right], \nonumber \\
\label{eq:Chern}
\end{eqnarray} 
where $\hat{R}(\bm{k}) = \bm{R}(\bm{k})/| \bm{R}(\bm{k})|$.

In Fig.~\ref{fig:pd_qwz}, we depict the phase diagram in $\alpha_3$-$m_0$ space.
We set the other parameters as 
$t_1=-0.3$, $t_2  =-0.3$ and $\alpha_1 = \alpha_2 =0.5$ for the panel (a) and 
$t_1=-0.5$, $t_2  =-1$ and $\alpha_1 = \alpha_2 =0.5$ for the panel (b),
for concreteness of the following discussions.
Here, the Chern number has been numerically computed~\cite{Fukui2005}.
There appear four phases, TI, normal insulator (NI), TSM, and normal semimetals (NSM). 
Here, when a Chern number is vanishing, we have said that the phase is normal, 
otherwise it is topological. 
Additionally, there exist the band touching lines (BTL) 
where the two bands touch at some momenta. The lines are represented by 
the black lines, $m_0=-2$, $0$, $2$, and $\alpha_3=0$.   

\subsection{Edge modes}
We now turn to the exact solution of the edge modes, under the open boundary condition in the $x$ direction.
The details of the method are presented in Appendix~\ref{sec:TM_method},
and we use the notation used there. 

The exact form of the dispersion relation is presented in Appendix~\ref{sec:disp_exact_qwz}.
To check whether the bulk-edge correspondence holds, we discuss the conditions for 
the existence of the edge solutions.  
Let us focus on the left edge modes. 
The existence of the edge solution can be examined by 
analyzing the eigenvalues of the transfer matrix $T$, i.e., $\lambda_1$ and $\lambda_2$ given by Eqs.~(\ref{eq:lambda_1}) and (\ref{eq:lambda_2}), respectively. 
They must satisfy $|\lambda_1| < 1$ and $|\lambda_2| < 1$ simultaneously.
These two conditions for the two eigenvalues guarantee that the edge solutions will decay exponentially 
in the bulk region. 
\begin{figure}[tb]
\begin{center}
\includegraphics[clip,width = 0.9\linewidth]{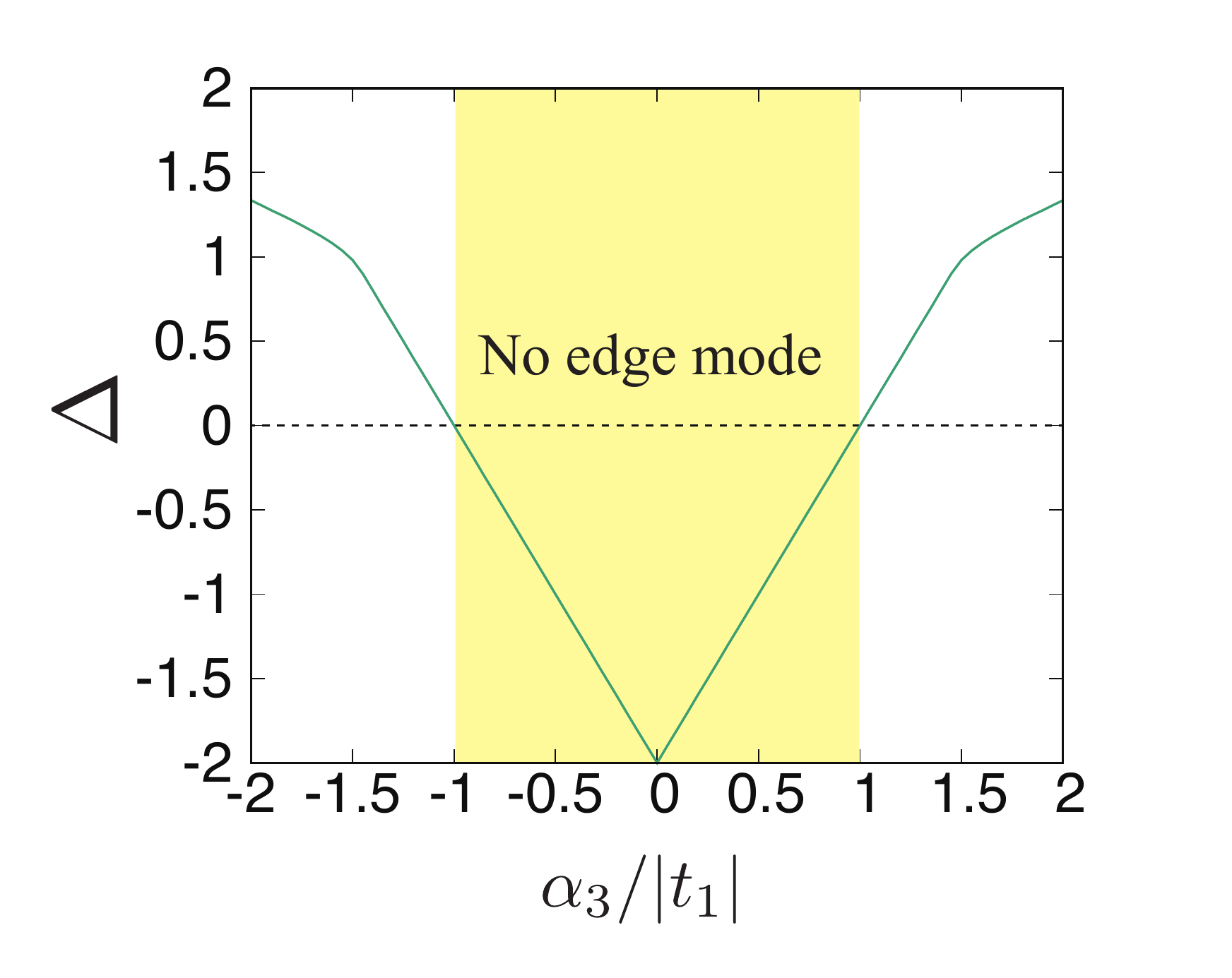}
\caption{The gap $\Delta$ of Eq.~(\ref{eq:delta_qwz}) as a function of $\alpha_3/|t_1|$.
The other parameters are set as $(t_1,t_2,\alpha_1,\alpha_2,m_0) = (-0.5,-1,0.5,0.5,-1.5)$.
The black dashed line denotes $\Delta = 0$.}
  \label{fig:delta_qwz}
 \end{center}
\end{figure}

In the present model, the matrices $A$ and $B$,
of which the transfer matrix $T$ consists [Eq.~(\ref{eq:tm_def})], are given by
\begin{eqnarray}
A= \begin{pmatrix}
t_1+\alpha_3 & -i\alpha_1\\
-i\alpha_1 & t_1 -\alpha_3 \\
\end{pmatrix}, \label{eq:QWZ_A}
\end{eqnarray}
and
\begin{widetext}
\begin{eqnarray}
B =\begin{pmatrix}
2 t_2 \sin k_y +2\alpha_3 (m_0+\cos k_y) & -2i \alpha_2 \sin k_y \\
2i \alpha_2 \sin k_y  & 2 t_2 \sin k_y  - 2\alpha_3 (m_0+\cos k_y) \\
\end{pmatrix}.
\nonumber \\
 \label{eq:QWZ_B}
\end{eqnarray}
\end{widetext}
We write $\mu$ for the eigenvalue of $A^{-1}A^\dagger$, and $\eta$ for that of $A^{-1}(EI_2 - B)$. 
As shown in Eqs.~(\ref{eq:lambda_1}) and (\ref{eq:lambda_2}), the two eigenvalues, $\lambda_1$ and $\lambda_2$, are 
written in terms of $\mu$ and $\eta$. 
We also have relations, $\mu=\lambda_1\lambda_2$ and $\eta=\lambda_1+\lambda_2$. 
Since $|\lambda_1|<1$ and $|\lambda_2|<1$, $\mu$ must satisfy $|\mu|=|\lambda_1\lambda_2|<1$.
From Eq.~(\ref{eq:QWZ_A}), one obtains 
\begin{eqnarray}
A^{-1}A^\dagger =\frac{1}{t_1^2-\alpha_3^2+\alpha_1^2}
\begin{pmatrix}
t_1^2-\alpha_3^2-\alpha_1^2  & 2i \alpha_1(t_1-\alpha_3) \\
2i \alpha_1(t_1+\alpha_3) & t_1^2-\alpha_3^2-\alpha_1^2 \\
\end{pmatrix}. \nonumber \\
\end{eqnarray}
Then, $\mu$ is given by 
\begin{eqnarray}
\mu_{\pm}=\frac{1}{t_1^2-\alpha_3^2+\alpha_1^2}\left[t_1^2-\alpha_3^2-\alpha_1^2\pm 2|\alpha_1|\sqrt{\alpha_3^2-t_1^2}\right]. \nonumber \\ \label{eq:mu_qwz}
\end{eqnarray}
From this Eq. (\ref{eq:mu_qwz}), one can easily show that, when $|t_1|>|\alpha_3|$, $|\mu|=1$. 
This implies no edge solutions for $|t_1|>|\alpha_3|$ from the above observations. 
Combining this fact and the phase diagram of Fig.~\ref{fig:pd_qwz},
we find that the emergence of the edge modes in the TSM phase depends on the parameter.
This means that the non-vanishing Chern number for the TSM is not sufficient for the existence of the edge modes. 
Therefore, the bulk-edge correspondence is broken down in the TSM phase,
because the TSM is realized even when $|t_1|>|\alpha_3|$. 
Meanwhile, the TI is realized only when $|t_1| < |\alpha_3|$ is satisfied, meaning the validity of the bulk-edge correspondence for the TI. 
\begin{figure}[t]
\begin{center}
\includegraphics[clip,width = 0.9\linewidth]{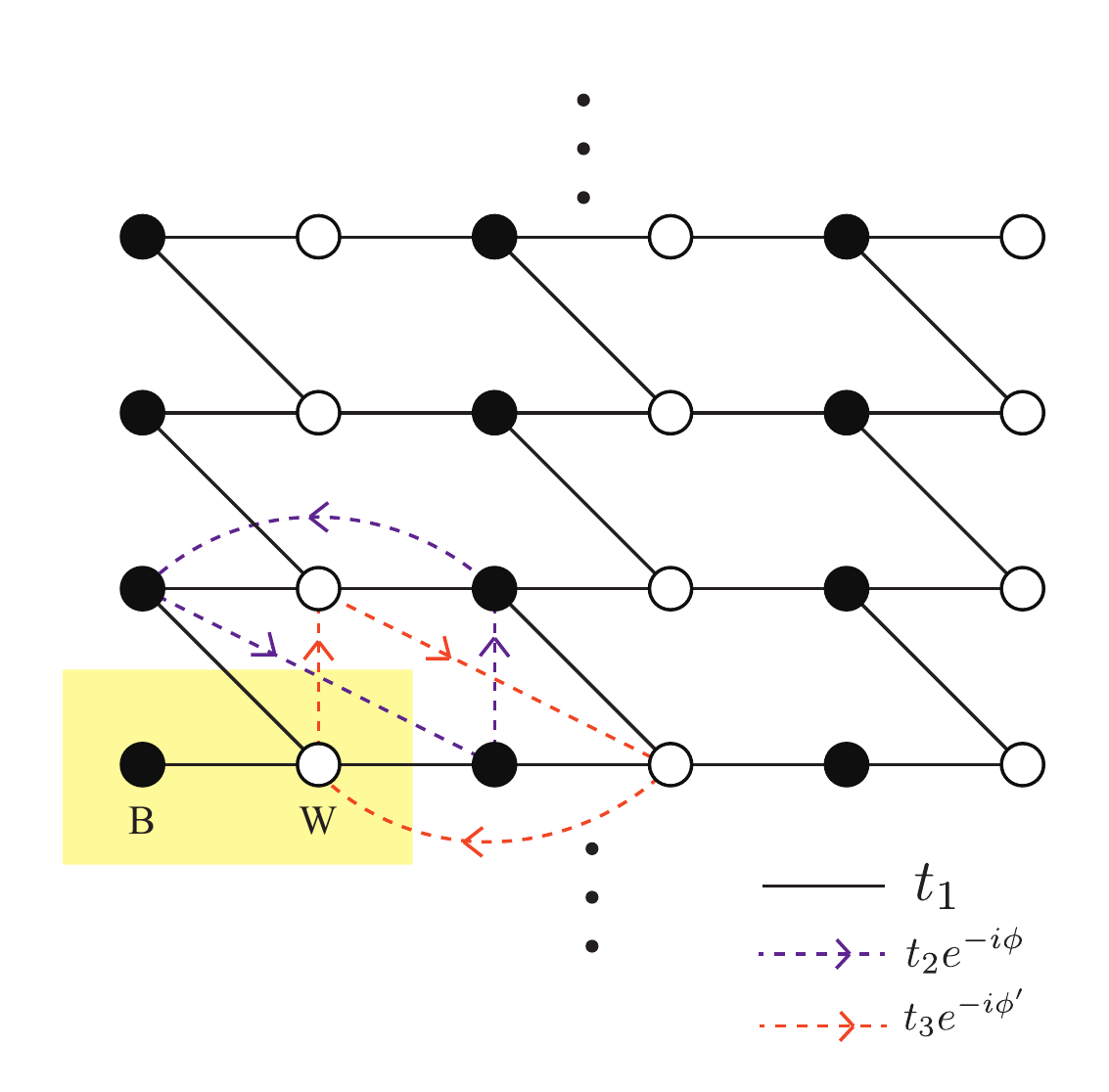}
\caption{Schematic figure of the Haldane-type model on a cylinder. 
The dots indicate the periodic boundary condition in the vertical direction.}
  \label{fig:1}
 \end{center}
\end{figure}

To ensure that the above discussion holds, 
we plot the dispersion under the open (periodic) boundary condition in the $x$ ($y$) direction
on Fig.~\ref{fig:band_qwz}.
Here we set the parameters such that the TSM is realized, and we examine the case with
$|\alpha_3|/|t_1| > 1$ [Fig.~\ref{fig:band_qwz}(a)] and $|\alpha_3|/|t_1| < 1$ [Fig.~\ref{fig:band_qwz}(b)].
The red lines are obtained by numerical diagonalization for a finite size system, 
and the blue and green lines are the exact results for the left and right edge modes, respectively.
Clearly, the former has the edge solutions with antichiral dispersion around $k_y=0$, while the latter does not. 
This result coincides with the above discussion.

To better understand the situation of the breakdown, 
we examine the relation between the overlapping of the two energy bands and 
the disappearance of the edge modes. 
From Figs.~\ref{fig:band_qwz}(a) and \ref{fig:band_qwz}(b), we can see the following: 
The edge modes always exist when the upper and lower bands are separated by the forbidden region, 
whereas they disappear immediately when the two bands touch at the single point $k_y=0$. 
These numerical computations show that the separation of the two bands is essential for 
the existence of the edge modes. 
To confirm this expectation, we want to introduce 
the separation parameter $\Delta$ between the upper and lower bands. 
As is well known, 
an essential spectrum of self-adjoint operators is invariant under compact perturbations, 
hence the energy spectrum of the extended states does not depend on the boundary 
condition in the $x$ direction in the present systems. 
By relying on this fact, we can use 
the energy spectrum of the bands, which is obtained under the periodic boundary condition 
in both the $x$ and $y$ directions, 
instead of that under the open boundary condition in the $x$ direction. 
We define the separation parameter $\Delta$ between the upper and lower bands by
\begin{eqnarray}
\Delta:=& \mathrm{min}_{k_y} \{ \mathrm{min} \{E_+(k_x,k_y)| k_x \in [-\pi,\pi]\}   \nonumber \\
-& \mathrm{max} \{E_-(k_x,k_y)| k_x \in [-\pi,\pi]\} \}.
\label{eq:delta_qwz}
\end{eqnarray}
When $\Delta$ is strictly positive, the two bands are separated 
by the forbidden region [cf. Fig.~\ref{fig:band_qwz}(a)]. 
Meanwhile, $\Delta $ can take strictly negative values. 
Then, the two bands significantly overlap 
on some range of $k_y$ [cf. Fig.~\ref{fig:band_qwz}(b)]. 
Clearly, the negative values of $\Delta$ have no meaning for the open boundary condition in the $x$ direction.

In the present QWZ-type model, the minimum value $\Delta$
is realized at $k_y=0$. 
For $\Delta= 0$, the two bands touch at the single point $k_y=0$. 
We plot $\Delta$ as a function of $\alpha_3/|t_1|$ in Fig.~\ref{fig:delta_qwz}. 
(For the other parameters, see the caption of Fig.~\ref{fig:delta_qwz}.) 
We indeed see that $\Delta$ is strictly positive (i.e., the two bands are separated by the forbidden region) 
for$ |\alpha_3|/|t_1|>1$, where the edge modes exist. Meanwhile, the two bands overlap for $|\alpha_3|/|t_1|<1$, 
where the edge modes are absent. Thus, the breakdown of the bulk-edge correspondence occurs  
when the upper and lower bands overlap on some range of $k_y$ in the TSM.

\section{Result 2: Haldane-type model \label{sec:GH}} 
We next study the Haldane-type model~\cite{Haldane1988}. 
The Hamiltonian on a cylinder geometry with the zigzag edge reads
\begin{eqnarray}
H_{\rm Haldane} =H _{\rm NN} + H_{\rm NNN}, \label{eq:ham_genHal}
\end{eqnarray}
where 
\begin{widetext}
\begin{eqnarray}
H_{\rm NN} &=& t_1  \sum_{\ell  =1}^{L_x} \sum_{m=1}^{L_y} 
\left[c^{\dagger}_{(\ell,m),W} c_{(\ell,m),B} + c^{\dagger}_{(\ell,m),W} c_{(\ell,m + 1),B}
+ (\mathrm{h.c.}) \right]
+  t_1 \sum_{\ell  =1}^{L_x-1} \sum_{m=1}^{L_y} 
\left[ c^{\dagger}_{(\ell,m),W} c_{(\ell + 1,m),B} 
+ (\mathrm{h.c.}) \right],
\label{eq:Ham_NN} 
\end{eqnarray}
and 
\begin{eqnarray}
H_{\rm NNN} &=&  \sum_{\ell  =1}^{L_x} \sum_{m=1}^{L_y} 
 \left[t_2e^{i\phi} c^{\dagger}_{(\ell,m),B} c_{(\ell,m + 1),B} +t_3 e^{i\phi^\prime}   c^{\dagger}_{(\ell,m),W} c_{(\ell,m + 1),W} \right]\nonumber \\
&+& (\mathrm{h.c.}) \nonumber \\
&+& \sum_{\ell  =1}^{L_x-1} \sum_{m=1}^{L_y} 
  \left[ t_2 e^{i\phi} c^{\dagger}_{(\ell + 1,m),B} c_{(\ell ,m),B} 
+ t_3 e^{i\phi^\prime}c^{\dagger}_{(\ell + 1,m),W} c_{(\ell,m),W}\right]\nonumber \\
&+& (\mathrm{h.c.}) \nonumber \\
&+& \sum_{\ell  =1}^{L_x-1} \sum_{m=1}^{L_y} 
\left[t_2 e^{i\phi}   c^{\dagger}_{(\ell,m ),B} c_{(\ell + 1,m - 1),B} 
+ t_3 e^{i\phi^\prime}c^{\dagger}_{(\ell ,m),W} c_{(\ell + 1,m -1),W}
\right] \nonumber \\ 
&+& (\mathrm{h.c.}).  \label{eq:Ham_NNN}
\end{eqnarray}
\end{widetext}
Here the subscripts, $B$ and $W$, denote the sublattice degrees of freedoms, as shown in Fig.~\ref{fig:1}.
The parameters $t_1$, $t_2$,  and $t_3$ are transfer integrals taken to be real,
and $\phi$ and $\phi^\prime$ are the phase factors for the next-nearest-neighbor hoppings. 
The model is the same as the Haldane's original model when $t_2 = t_3$ and $\phi^\prime = -\phi$. 
As is well known, the chiral edge modes appear since the bulk Chern number takes $\pm 1$, manifesting the topologically nontrivial nature~\cite{Haldane1988}. 
On the other hand, it was pointed out in Ref.~\onlinecite{Colomes2018} that the antichiral edge modes appear when 
$t_2 = t_3$ and $\phi^\prime =  \phi$.

\subsection{Bulk properties}
Similarly to the QWZ-type model, we first summarize the bulk properties.
Imposing the periodic boundary condition in both $x$ and $y$ directions 
and performing the Fourier transformation, 
the Hamiltonian of Eq. (\ref{eq:ham_genHal}) can be written as
\begin{eqnarray}
H = \sum_{(k_x,k_y)}
\bm{\Psi}^\dagger(k_x,k_y) 
\mathcal{H}(k_x,k_y)
\bm{\Psi}(k_x,k_y), \nonumber \\
\end{eqnarray}
where $\bm{\Psi}(k_x,k_y) = \left(c_{(k_x,k_y),B}, c_{(k_x,k_y),W}\right)^{\rm T}$, and
$\mathcal{H}(k_x,k_y)$ is the $2\times 2$ matrix written as
\begin{eqnarray}
\mathcal{H}(k_x,k_y) = R_0(k_x,k_y) I_2 + \bm{R}(k_x,k_y) \cdot \bm{\tau}
\end{eqnarray}
with
\begin{subequations}
\begin{eqnarray}
R_0(k_x,k_y) =& t_2 \mathrm{Re} \left[ e^{i \left(\phi - k_x \right)} + e^{i\left(\phi + k_y\right)}  + e^{i\left(\phi + k_x-k_y\right)} \right] \nonumber \\
+& t_3 \mathrm{Re} \left[ e^{i\left(\phi^\prime - k_x\right)} + e^{i\left(\phi^\prime + k_y\right)}  + e^{i\left(\phi^\prime + k_x-k_y\right)}   \right], \nonumber \\
\end{eqnarray}
\begin{eqnarray}
R_1(k_x,k_y) = t_1 \left(1 + \cos k_x +\cos k_y \right),
\end{eqnarray}
\begin{eqnarray}
R_2 (k_x,k_y) = t_1 \left( \sin k_x +\sin k_y \right),
\end{eqnarray}
and 
\begin{eqnarray}
R_3(k_x,k_y) = &t_2 \mathrm{Re} \left[ e^{i\left(\phi - k_x\right)} + e^{i\left(\phi + k_y\right)}  + e^{i\left(\phi + k_x-k_y\right)}\right]  \nonumber \\
- &t_3 \mathrm{Re} \left[e^{i\left(\phi^\prime - k_x\right)} + e^{i\left(\phi^\prime + k_y\right)}  + e^{i\left(\phi^\prime + k_x-k_y\right)}  \right]. \nonumber \\
\end{eqnarray}
\end{subequations}

\begin{figure}[t]
\begin{center}
\includegraphics[clip,width = \linewidth]{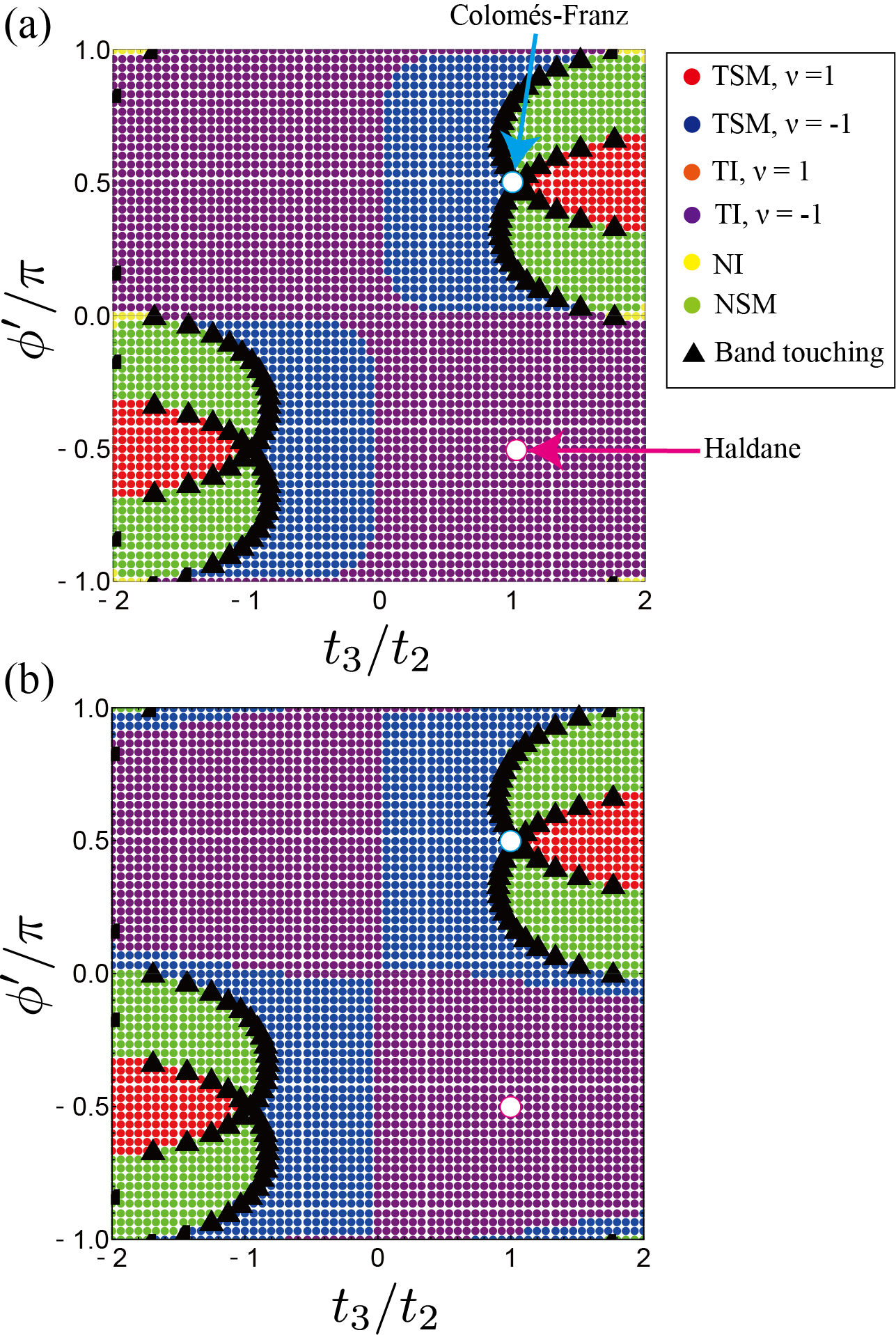}
\caption{The phase diagram of the Haldane-type model for 
(a) $t_1=1$, $t_2=0.2$, and $\phi = \frac{\pi}{2}$ and (b) $t_1=1$, $t_2=0.5$, and $\phi = \frac{\pi}{2}$.
Black triangles represent the band touching cases where $|\bm{R}|$ becomes 0 in some $\bm{k}$, 
i.e., the solutions of either Eq.~(\ref{eq:gapclose_1}) or Eq.~(\ref{eq:gapclose_2}),
obtained numerically.}
  \label{fig:pd}
 \end{center}
\end{figure}
\begin{figure*}[t]
\begin{center}
\includegraphics[clip,width = \linewidth]{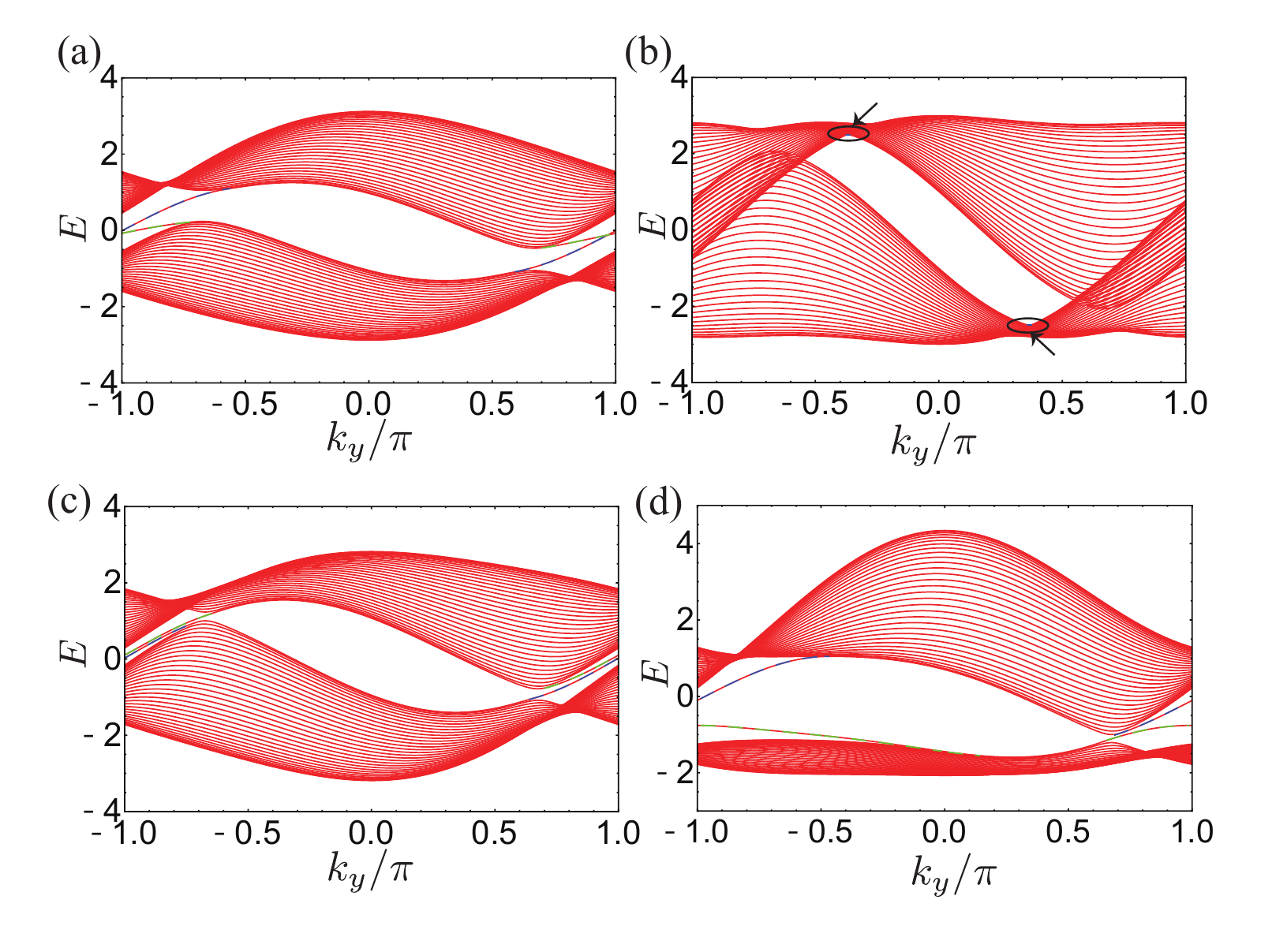}
\caption{
The band structures for the Haldane-type model on a cylinder
with $(t_1 , t_2, t_3, \phi,\phi^\prime) = $ 
(a) $\left(1, 0.2, 0.08, \frac{\pi}{2}, \frac{\pi}{3} \right)$,
(b) $\left(1, 0.5, 0.4, \frac{\pi}{2}, \frac{\pi}{2} \right)$,
(c) $\left(1, 0.2, 0.2 , \frac{\pi}{2}, \frac{3 \pi}{5} \right)$,
and (d) $\left(1, 0.2, 0.38, \frac{\pi}{2}, 0 \right)$. 
Note that the panels (a) and (b) are for the TSM, (c) is for the normal semimetal, and (d) is for the normal insulator. 
Red solid lines are obtained by the numerical diagonalization of $\mathcal{H}(k_y)$ for $L_x=32$,
and blue and green broken lines are the exact solutions for the left and right edge modes, respectively.
The black ellipses in the panel (b) highlight the regions where the left edge mode exists.}
\label{fig:generic}
\end{center}
\vspace{-10pt}
\end{figure*}
The band touching between the conduction and valence bands occurs when 
$|\bm{R}(k_x,k_y)| =0$ is satisfied at some momenta. 
In fact, $R_1 (k_x,k_y)  = R_2 (k_x,k_y)= 0$ is satisfied 
at two momenta, $ \bm{k}_1 := \left(- \frac{2\pi}{3},\frac{2\pi}{3} \right)$ 
and $\bm{k}_2 := \left(\frac{2\pi}{3}, -\frac{2\pi}{3} \right)$, 
which correspond to K$^\prime$ and K points in the conventional 
notation for the high symmetry points in the first Brillouin zone of the honeycomb lattices, respectively.
Thus, the condition of the band touching is dictated by $R_3(\bm{k}_{1}) = 0$ or $R_3(\bm{k}_{2}) = 0$;
the former is written as
\begin{eqnarray}
-\frac{3}{2} \left(t_2 \cos\phi - t_3 \cos \phi^\prime \right) -\frac{3\sqrt{3}}{2} \left(t_2 \sin \phi - t_3 \sin \phi^\prime \right)=0,  \nonumber \\
\label{eq:gapclose_1}
\end{eqnarray}
while the latter as 
\begin{eqnarray}
-\frac{3}{2} \left(t_2 \cos\phi - t_3 \cos \phi^\prime \right) + \frac{3\sqrt{3} }{2} \left(t_2 \sin \phi - t_3 \sin \phi^\prime \right)=0. 
 \nonumber \\
\label{eq:gapclose_2}
\end{eqnarray}
If the band touching does not occur at any momenta, the Chern number of Eq.~(\ref{eq:Chern}) is well-defined.

In Fig.~\ref{fig:pd}, we draw $t_3/t_2$-$\phi^\prime$ phase diagrams of the present model,
fixing $t_2 = 0.2$, $\phi = \frac{\pi}{2}$ for Fig.~\ref{fig:pd}(a), and $t_2 = 0.5$, $\phi = \frac{\pi}{2}$ for Fig.~\ref{fig:pd}(b).
Note that the band touching curves, given by the solutions of either (\ref{eq:gapclose_1}) or (\ref{eq:gapclose_2}), 
are denoted by black triangles.
We find that, for both Figs.~\ref{fig:pd}(a) and \ref{fig:pd}(b), 
tthe TSM phases appear on the parameter space. 
\begin{figure}[tb]
\begin{center}
\includegraphics[clip,width = 0.9\linewidth]{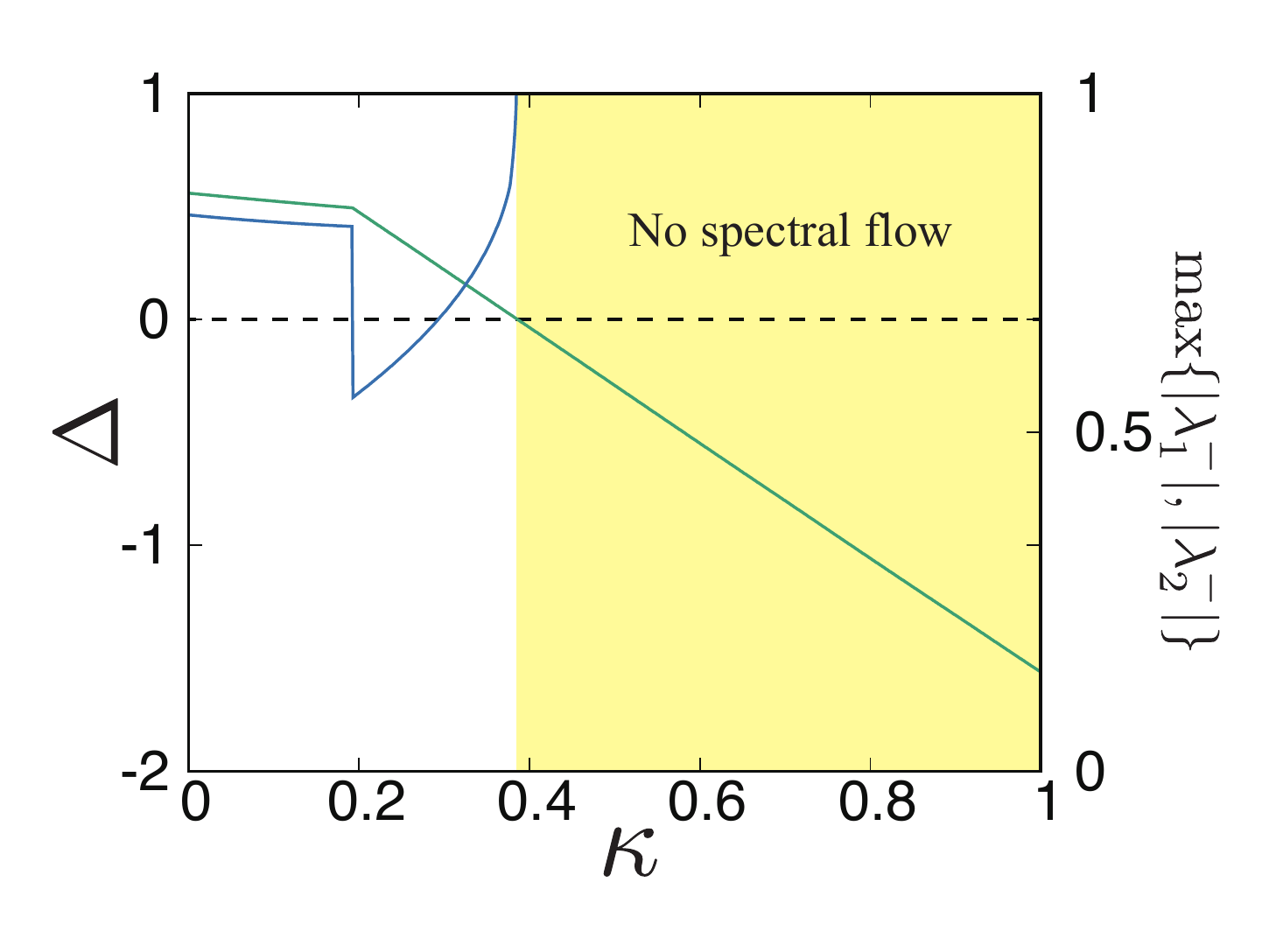}
\caption{The green line is $\Delta$ of Eq.~(\ref{eq:delta_qwz}) (the left vertical axis).
The blue line is the maximum of $\{|\lambda_1^-|,|\lambda_2^- | \}$ (the right vertical axis).
The black dashed line denotes $\Delta = 0$ for the left vertical axis.
The jump of the blue line at  $\kappa \sim 0.2$ is due to the difference between the two values of $k_y^{\rm min}$ 
which give the minimum values $\Delta$ of the band separation.
Since the present transfer matrix method is not applicable to extended states, 
we do not plot $\lambda$ for $k_y^{\rm min}$ which gives an extended state in the yellow region.}
  \label{fig:delta_haldane}
 \end{center}
\end{figure}

\subsection{Edge modes}
Let us turn to the analysis of the edge modes. 
The exact dispersion relation is obtained for any parameters, and we present it in Appendix~\ref{sec:disp_exact_hal}.
Here, we show some results in Fig.~\ref{fig:generic}; we plot the dispersion relation
of the Hamiltonian of Eq.~(\ref{eq:ham_genHal}) obtained by 
the numerical diagonalization (red lines) and 
the exact solutions for the left (blue broken lines) and right (green broken lines) edge modes 
with some representative values of the parameters. 

We can see that the antichiral edge modes exist in Fig.~\ref{fig:generic}(a), 
while they do not in Fig.~\ref{fig:generic}(b). 
Strictly speaking, in Fig.~\ref{fig:generic}(b), 
there are very tiny regions ($k_y \sim \pm 0.36\pi$, highlighted by the black ellipses) 
where only the left edge modes survive. 
However, the tiny edge modes does not an antichiral dispersion, nor do they exhibit the spectral flow. 
Here, the spectral flow means that an edge mode energetically connects conduction and valence bands.
In this sense, these results indicate that the breakdown of the bulk-edge correspondence
occurs also in the Haldane-type model. 
Although we want to discuss the relation between 
the breakdown and the overlapping of the bands, we have been unable to find a simple criterion 
for the existence of the antichiral edge modes in the Haldane-type model, 
in contrast to the QWZ-type model. 
Of course, Figs.~\ref{fig:generic}(a) and \ref{fig:generic}(b) 
also show that the spectral flow is abruptly broken at the boundary of some parameter region 
which contains the values of the parameters in Fig.~\ref{fig:generic}(b). 
To investigate the behavior of 
$\Delta$ of Eq.~(\ref{eq:delta_qwz}) near the critical region, 
we introduce a family of the parameter sets $t_1(\kappa), t_2(\kappa), t_3(\kappa), \phi(\kappa), \phi' (\kappa)$
with an additional parameter $\kappa \in [0,1]$ 
that linearly interpolates between the two sets  $ t_1, t_2, t_3, \phi, \phi^\prime$ in
Figs.~\ref{fig:generic}(a) and \ref{fig:generic}(b).  
Clearly, a value of $\kappa$ hits a set of the critical values of the parameters.  
To check the existence of the edge mode, we calculate the eigenvalue $\lambda(k_y^{\rm min})$ of the transfer matrix
[see Eq.~(\ref{eq:eigen_tm}) for its definition]. 
Here, $k_y^{\rm min}$ stands for the momentum $k_y$ for which 
the minimum value $\Delta$ is realized. In the following, we will treat only the left edge mode, 
and we write $\lambda=\lambda(k_y^{\rm min})$, dropping the $k_y^{\rm min}$ dependence of $\lambda(k_y^{\rm min})$.

We recall the following:  
(i) If the absolute values of two eigenvalues, $\lambda_1$ and $\lambda_2$, of the transfer matrix are both smaller than 1, 
then the left edge mode exists. 
Otherwise, it does not exist. 
(ii) There are two types of pairs, $\{\lambda_1^+, \lambda_2^+ \}$ and $\{ \lambda_1^-, \lambda_2^- \}$, of 
the eigenvalues. (see Appendix.~\ref{sec:disp_exact_hal} for the details.)  
However, $\{\lambda_1^+, \lambda_2^+ \}$ does not satisfy 
the above condition (i). Therefore, it is enough to consider $\{ \lambda_1^-, \lambda_2^- \}$. 
Figure~\ref{fig:delta_haldane} shows the $\kappa$ dependence of 
$\Delta$, and that of $\mathrm{max} \{ |\lambda_1^-|, |\lambda_2^-| \}$. 
We see that $\Delta$ is positive (i.e., the two bands do not overlap) for $\kappa \lesssim 0.4$, 
where $\mathrm{max} \{ |\lambda_1^-|, |\lambda_2^-| \}$  is smaller than 1, and thus the edge mode exists. 
On the other hand, for $\kappa \sim 0.4$, the band overlapping occurs,
and the edge mode disappears at $k_y^{\rm min}$ since one of $|\lambda_1^-|$ and $|\lambda_2^-|$
is greater than or equal to 1. 
(These values of $|\lambda_1^-|$ and $|\lambda_2^-|$ have no meaning  
because the present transfer matrix method is not applicable to extended states.)
This implies that the spectral flow is broken near $k_y^{\rm min}$. 
Therefore, the critical value of $\kappa$ at which the spectral flow is abruptly broken
coincides with the other critical value at which the two bands touch at 
the single point $k_y^{\rm min}$ within the numerical precision. 
We also stress that in the case of the Haldane-type model, 
the edge modes do not necessarily disappear even beyond the transition point, 
in contrast to the QWZ-type model. 
However, the spectral flow from the lower band to the upper band disappears similarly to the QWZ-type model.

Next, we discuss the nature of the edge modes.
In Fig.~\ref{fig:generic}(a), 
we see that the left and right edge modes have the same sign of the group velocities, meaning that the antichiral edge modes are realized. 
Interestingly, they also appear in the normal semimetal
as shown in Fig.~\ref{fig:generic}(c).
Moreover, Fig.~\ref{fig:generic}(d) shows that the edge modes in the trivial insulator exhibit 
an antichiral-like dispersion around $k_y =  \pi$.
Nevertheless, there is a crucial difference between trivial and nontrivial cases, 
i.e., the spectral flow occurs in Fig.~\ref{fig:generic}(a), while it does not in Figs.~\ref{fig:generic}(c) and \ref{fig:generic}(d). 
Additionally, for the band touching case 
that we will address in Fig~\ref{fig:exact_HCF} of Appendix~\ref{sec:disp_exact_2}, 
the edge modes connect two gapless points in the bulk, which is reminiscent of graphene under the zigzag edge. 

\section{Summary \label{sec:summary}} 
We have investigated the bulk-edge correspondence in topological semimetals (TSMs) 
in which the Chern number can be defined by using the valence band structure because the valence and 
conduction bands are separated by the forbidden region on the momentum space. 
By relying on the transfer matrix method, we have derived the exact solutions of the antichiral edge modes 
for the QWZ-type model and the Haldane-type model. 
In both of the two models, 
we have found that the non-vanishing Chern number does not necessarily lead to the emergence of the antichiral edge 
modes. In fact, it depends on the model parameters, meaning that the bulk-edge correspondence is broken down 
in these TSMs.  
The intuitive explanation of the bulk-edge correspondence is as follows: 
A non-vanishing Chern number is a consequence of a non-trivial topological structure 
of the wavefunctions of the valence band. 
Therefore, one can expect that, when an open boundary condition is imposed, the wavefunctions at the edges of the sample 
exhibit a peculiar behavior. This is nothing but the emergence of chiral or antichiral edge modes. 
However, surprisingly, there appears to be no antichiral edge mode for some parameters although the Chern number is non-vanishing.
We have also found that the vanishing of the edge modes is associated with 
the overlapping between the upper and lower bands in TSMs. 
Actually, the overlapping obstructs the spectral flow from the lower band to 
the upper band such that the flow does not touch neither of the bands along the way. 
On the other hand, in the system with no open boundary, 
the Chern number is well-defined on the two-dimensional momentum space, 
where the upper and lower bands are still separated by some forbidden region. 
These give an explanation of the breakdown of the bulk-edge correspondence.

As for the Haldane-type model, we have found that the antichiral edge modes can appear in not only 
the TSM phases but also the trivial phases. 
However, it is only in the TSM phases that the antichiral edge modes exhibit the spectral flow which connects the valence and conduction bands. 

\acknowledgements
The authors are grateful to the anonymous referee for the helpful comment on the relation between the vanishing of the edge modes and the overlapping between the bands.
T. M. is supported by the JSPS KAKENHI, Grants No.~JP17H06138 and No.~JP20K14371, 
Japan.

\appendix
\section{Review of the method to obtain exact edge solutions \label{sec:TM_method}} 
In this appendix, we review the transfer matrix method~\cite{Mizoguchi2020} to obtain exact solutions of 
edge modes. 
As we have emphasized, this method is applicable to generic lattice models
whose transfer matrix is given as a $4\times 4$ matrix.

For concreteness, let us focus on the two-dimensional models. 
Namely, we consider a lattice model on a cylinder, 
where an open boundary condition and a periodic boundary condition are imposed, 
in the $x$ and $y$ directions, respectively. 
Annihilation operators of fermions on each unit cell are expressed by a two-component vector:
\begin{eqnarray}
\bm{\Psi}_{(\ell,m)} = 
\left(
\begin{array}{c}
c_{(\ell,m),1} \\
c_{(\ell,m),2} \\
\end{array}
\right),
\end{eqnarray}
where the indices $1$ and $2$ denote internal degrees of freedom. 

Consider a tight-binding Hamiltonian:
\begin{eqnarray}
H = \sum_{\ell, \ell^\prime= 1}^{L_x} \sum_{m, m^\prime = 1}^{L_y}
\bm{\Psi}^\dagger_{(\ell,m)} \hat{h}_{(\ell,m), (\ell^\prime,m^\prime)}\bm{\Psi}_{(\ell^\prime,m^\prime)}, \label{eq:ham_general}
\end{eqnarray}
where $\hat{h}_{(\ell,m), (\ell^\prime,m^\prime)}$ is a $2 \times 2$ matrix 
satisfying 
\begin{eqnarray}
\hat{h}_{(\ell,m), (\ell^\prime,m^\prime)} = 0, \hspace{1mm} \mathrm{if} \hspace{1mm} |\ell -\ell^\prime| \geq 2,
\end{eqnarray}
and $L_x$ ($L_y$) is the number of unit cells along the $x$ ($y$) direction.

Due to the periodic boundary condition in the $y$ direction, we can perform the Fourier transformation in the $y$ direction: 
\begin{eqnarray}
c_{\ell,k_y,\alpha} = \frac{1}{\sqrt{L_y}} \sum_{m=1}^{L_y} e^{-i k_y m} c_{(\ell,m),\alpha} \label{eq:Fourier}
\end{eqnarray}
with $\alpha = 1,2$.
The inverse Fourier transformation is 
\begin{eqnarray}
c_{(\ell,m),\alpha} = \frac{1}{\sqrt{L_y}} \sum_{k_y} e^{i k_y m} c_{\ell,k_y,\alpha}. \label{eq:inv_Fourier}
\end{eqnarray}
Substituting Eq. (\ref{eq:inv_Fourier}) into Eq. (\ref{eq:ham_general}), we have  
\begin{eqnarray}
H = \sum_{k_y} \sum_{\ell, \ell^\prime} \bm{\Psi}_{\ell}^\dagger(k_y)[\mathcal{H} (k_y)]_{\ell, \ell^\prime} \bm{\Psi}_{\ell^\prime} (k_y),
\end{eqnarray}
where
\begin{eqnarray}
\bm{\Psi}_{\ell} (k_y) = 
\left(
\begin{array}{c}
c_{\ell,k_y, 1} \\
c_{\ell,k_y, 2}\\
\end{array}
\right),
\end{eqnarray}
and $[\mathcal{H} (k_y)]_{\ell, \ell^\prime}$ is a $2\times 2$ matrix that can be written in general as
\begin{eqnarray}
[\mathcal{H} (k_y)]_{\ell, \ell^\prime}
 = B (k_y) \delta_{\ell, \ell^\prime} + A(k_y) \delta_{\ell, \ell^\prime-1} + A^\dagger(k_y)\delta_{\ell, \ell^\prime+1}. \nonumber \\
  \label{eq:ham}
\end{eqnarray}

Let $[a^{\rm L}(k_y)]^\dagger$ be the left edge mode which is written as 
\begin{eqnarray}
[a^{\rm L}(k_y)]^\dagger = \sum_{\ell = 1}^{L_x}  \bm{\Psi}^\dagger_{\ell}(k_y) \cdot \bm{\varphi}_{\ell} (k_y) \label{eq:eigen}
\end{eqnarray}
with the amplitudes $\bm{\varphi}_\ell(k_y)$. 
This operator $[a^{\rm L}(k_y)]^\dagger$ satisfies the commutation relation 
\begin{eqnarray}
\left[H,[a^{\rm L}(k_y)]^\dagger \right] = E  (k_y)[a^{\rm L}(k_y)]^\dagger, \label{eq:comm}
\end{eqnarray} 
where $E  (k_y)$ is the eigenenergy of the edge mode.
Then, substituting Eqs.~(\ref{eq:ham}) and (\ref{eq:eigen}) into Eq.~(\ref{eq:comm}), we have
\begin{eqnarray}
 &&A(k_y) \bm{\varphi}_{\ell  + 1} (k_y)+ A^\dagger(k_y)\bm{\varphi}_{\ell  - 1} (k_y)+ B(k_y) \bm{\varphi}_{\ell} (k_y) \nonumber \\
  &=&E  (k_y) \bm{\varphi}_{\ell} (k_y). \nonumber \\ 
  \label{eq:schroedinger}
\end{eqnarray}
In the following, we assume $\mathrm{det} A(k_y) \neq 0$. 
When $\mathrm{det} A(k_y) =0$, we choose the model parameters 
to satisfy $\mathrm{det} A(k_y) \neq 0$, and we take the limit $\mathrm{det} A(k_y) \rightarrow 0$ 
by varying the parameters, after all the calculations are done. 
Equation (\ref{eq:schroedinger}) can also be written as 
\begin{eqnarray}
\left(
\begin{array}{c}
\bm{\varphi}_{\ell + 1} (k_y) \\
\bm{\varphi}_{\ell} (k_y)\\
\end{array}
\right)
= T(k_y)
\left(
\begin{array}{c}
\bm{\varphi}_{\ell } (k_y) \\
\bm{\varphi}_{\ell-1} (k_y)\\
\end{array}
\right), \label{eq:tm}
\end{eqnarray}
where the transfer matrix $T(k_y)$ is given by
\begin{eqnarray}
T(k_y) = \left( 
\begin{array}{cc}
A^{-1} (k_y)[E  (k_y) I_{2} -B(k_y)] &  - A^{-1} (k_y) A^{\dagger}(k_y)\\
I_{2}  & 0\\
\end{array} 
\right), \nonumber \\ \label{eq:tm_def}
\end{eqnarray} 
with the $2\times 2$ identity matrix $I_2$.

In the following, we abbreviate $T(k_y)$ as $T$, similarly $A=A(k_y)$, $B=B(k_y)$, and $E=E(k_y)$.
To obtain the solution of $\bm{\varphi}_{\ell}$, let us consider the eigenvalue problem of $T$.
We write $(\bm{\psi}_1,\bm{\psi}_2)^{\rm T}$ for the eigenvector of the transfer matrix $T$, and $\lambda$ for 
the eigenvalue. 
Here, $\bm{\psi}_1$ and $\bm{\psi}_2$ are two-component vectors to be determined.  
Then, the eigenvalue equation can be written as 
\begin{eqnarray}
T \left(
\begin{array}{c} \bm{\psi}_1\\ \bm{\psi}_2 \\ \end{array} 
\right)
&=&\left( \begin{array}{c}A^{-1}[E I_{2} -B] \bm{\psi}_1 - A^{-1}  A^{\dagger} \bm{\psi}_2 \\ 
\bm{\psi}_1 \\ \end{array} \right)  \nonumber \\
&=&\lambda \left(
\begin{array}{c} \bm{\psi}_1\\ \bm{\psi}_2 \\ \end{array} 
\right), \nonumber \\
\label{eq:eigen_tm}
\end{eqnarray}
We will construct the solutions of (\ref{eq:tm}) by using the eigenvectors of $T$. 
From the second row of Eq.~(\ref{eq:eigen_tm}), we have $\bm{\psi}_1  =\lambda\bm{\psi}_{2}$. 
Substituting this into the first row, 
we have 
\begin{eqnarray}
A^{-1}[E  I_{2} -B] \lambda \bm{\psi}_{2}
- A^{-1} A^{\dagger}\bm{\psi}_{2}= \lambda^2\bm{\psi}_{2}.  \label{eq:eigenTM_2}
\end{eqnarray}
If $\lambda =0$, then one has $\bm{\psi}_1=0$ and $A^{-1}A^\dagger \bm{\psi}_2=0$. 
These imply $(\bm{\psi}_1, \bm{\psi}_2)^{\rm T}=0$. 
Thus, one has $\lambda\neq 0$.  
Using Eq.~(\ref{eq:eigenTM_2}) and its hermitian conjugate, 
it can be found that the four solutions of the eigenvalue equation of $T$ 
are given as~\cite{Mizoguchi2020,Molinari1997}
\begin{eqnarray}
\lambda_1, \frac{1}{\lambda_1^{\ast}}, \lambda_2, \frac{1}{\lambda_2^{\ast}}.  \label{eq:lambdas}
\end{eqnarray}
Let us consider the left edge mode for which the eigenvalue $\lambda$ must satisfy $0<|\lambda|<1$
because $\bm{\varphi}_\ell \rightarrow 0$ as $\ell \rightarrow \infty$. 
Therefore, from (\ref{eq:lambdas}), we want to find 
the two eigenvalues, $\lambda_1$ and $\lambda_2$, 
which satisfy $0<|\lambda_i|<1$, $i=1,2$. 
Namely, the left edge mode is constructed by the two eigenvectors of $T$ with 
the eigenvalues, $\lambda_1$ and $\lambda_2$. 

Using $\bm{\psi}_1=\lambda \bm{\psi}_2$, we write $(\lambda_j \bm{\chi}_j, \bm{\chi}_j)$ 
for the eigenvector of $T$ 
with the eigenvalue $\lambda_j$, $j=1,2$. 
Then, the left edge mode satisfying Eq.~(\ref{eq:tm}) can be written as
\begin{eqnarray}
\left( 
\begin{array}{c} 
\bm{\varphi}_{\ell} \\
\bm{\varphi}_{\ell-1} \\
\end{array} 
\right) 
= d_1 \lambda_1^{\ell-1}
\left( 
\begin{array}{c} 
\lambda_1 \bm{\chi}_1\\
\bm{\chi}_1 \\
\end{array} 
\right)
+ d_2 \lambda_2^{\ell-1}
\left( 
\begin{array}{c} 
\lambda_2 \bm{\chi}_2\\
\bm{\chi}_2 \\
\end{array} 
\right), \nonumber \\
\label{eq:solution_eigenvec}
\end{eqnarray}
where $d_1$ and $d_2$ are coefficients. 
These two coefficients can be determined by taking into account the 
Dirichlet boundary condition at the left edge, i.e., $\bm{\varphi}_{0} = 0$.
This leads to
\begin{eqnarray}
d_1 \bm{\chi}_1 + d_2 \bm{\chi}_2 = 0.
\end{eqnarray}
The key observation is that, to obtain the nontrivial solution (i.e., the solution other than $d_1 = d_2 =0$), 
two vectors $\bm{\chi}_1$ and $\bm{\chi}_2$ have to be parallel to each other,
i.e., $\bm{\chi}_1 = C \bm{\chi}_2$ holds with $C$ being the constant.
Setting $C=1$ and substituting this relation into Eq.~(\ref{eq:eigenTM_2}),
one has 
\begin{eqnarray}
A^{-1}(E  I_{2} -B)\lambda_1 \bm{\chi}_{1}
- A^{-1} A^{\dagger}\bm{\chi}_{1}= \lambda_1^2 \bm{\chi}_{1}, \label{eq:eigenTM3}
\end{eqnarray}
and 
\begin{eqnarray}
A^{-1}(E  I_{2} -B)\lambda_2 \bm{\chi}_{1}
- A^{-1} A^{\dagger}\bm{\chi}_{1}= \lambda_2^2 \bm{\chi}_{1}.\label{eq:eigenTM4}
\end{eqnarray}
By further subtracting (\ref{eq:eigenTM4}) from (\ref{eq:eigenTM3}), one has
\begin{eqnarray}
A^{-1}(E  I_{2} -B)(\lambda_1-\lambda_2)  \bm{\chi}_{1}
= (\lambda_1^2-\lambda_2^2) \bm{\chi}_{1}. \label{eq:tm5}
\end{eqnarray}
In the following, we assume $\lambda_1\neq \lambda_2$. 
If necessary, we take 
the limit $\lambda_1-\lambda_2 \rightarrow 0$ from $\lambda_1\neq \lambda_2$. 
Equation~(\ref{eq:tm5}) indicates that $\bm{\chi}_1$ is the eigenvector of $A^{-1}(E  I_{2} -B)$ with the eigenvalue 
$\lambda_1 + \lambda_2$,
if $\lambda_1 \neq \lambda_2$.
Further, combining this fact and (\ref{eq:eigenTM3}), we find that $\bm{\chi}_1$ is the eigenvector of $A^{-1}A^{\dagger}$ with the eigenvalue $\lambda_1 \lambda_2$.
This is crucial because $A^{-1}A^{\dagger}$ is a $2 \times 2$ matrix 
which does not contain the unknown energy eigenvalue $E$,
thus its eigenvalues and eigenvectors can always be obtained exactly, without assigning any special conditions. 

To proceed further with the analysis, let $\bm{u}_j$ ($j=1,2$) be the eigenvector of $A^{-1}A^{\dagger}$ whose eigenvalue is $\mu_j$. 
Then, the energy eigenvalue $E $ can be determined 
so that $\bm{u}_j$ becomes the eigenvector of $A^{-1}(E  I_{2} -B)$ as well.
This can be achieved by solving a linear equation with respect to $E $:
\begin{eqnarray}
[\bm{u}_j]_2 [A^{-1}(E  I_{2} -B) \bm{u}_j ]_1 = [\bm{u}_j]_1 [A^{-1}(E  I_{2} -B) \bm{u}_j ]_2,\nonumber \\ \label{eq:E_to_solve}
\end{eqnarray}
where $[\cdots]_l $ stands for the $l$-th component of the vector. 
Then, the remaining task is to check whether the decaying solution exists or not.
Let $\eta_j$ be the eigenvalue of $A^{-1}(E  I_{2} -B)$ whose eigenvector is $\bm{u}_j$.
Then, using $\eta_j = \lambda_1 + \lambda_2 $ and $\mu_j = \lambda_1 \lambda_2$,
we have 
\begin{subequations}
\begin{eqnarray}
\lambda^{j}_1 = \frac{\eta_j}{2} + \sqrt{\frac{\eta_j^2 }{4} -\mu_j}, \label{eq:lambda_1}
\end{eqnarray} 
and
\begin{eqnarray}
\lambda^{j}_2 = \frac{\eta_j}{2} - \sqrt{\frac{\eta_j^2 }{4} -\mu_j}.\label{eq:lambda_2}
\end{eqnarray} 
\end{subequations}
The conditions, $|\lambda_1|<1$ and $|\lambda_2|<1$, must be satisfied simultaneously 
for the existence of the solution. 

So far, we have explained the derivation of the left edge mode.
The same method can be applied to derive the right edge mode, 
as we outline below. 
Similarly to Eq.~(\ref{eq:tm}), the eigenvalue equation can be written as
\begin{eqnarray}
\left(
\begin{array}{c}
\bm{\varphi}_{\ell -1 }  \\
\bm{\varphi}_{\ell} \\
\end{array}
\right)
= \tilde{T}
\left(
\begin{array}{c}
\bm{\varphi}_{\ell}\\
\bm{\varphi}_{\ell + 1}\\
\end{array}
\right) \label{eq:tm_right}
\end{eqnarray}
with 
\begin{eqnarray}
\tilde{T} 
= \left(
\begin{array}{cc}
(A^{\dagger})^{-1} (EI_2-B) & -(A^{\dagger})^{-1}A \\
I_2 & 0 \\
\end{array}
\right).
\end{eqnarray}
Using the two eigenvectors $\tilde{\bm{\chi}}_j$ of $\tilde{T}$ with the eigenvalue $\tilde{\lambda}_j$, $j=1,2$, 
the right edge mode can be written as 
\begin{eqnarray}
\left( 
\begin{array}{c} 
\bm{\varphi}_{\ell-1} \\
\bm{\varphi}_{\ell} \\
\end{array} 
\right) 
= \tilde{d}_1 \tilde{\lambda}_1^{L_x-\ell + 1 }
\left( 
\begin{array}{c} 
\tilde{\lambda}_1 \tilde{\bm{\chi}}_1\\
\tilde{\bm{\chi}}_1 \\
\end{array} 
\right)
+ \tilde{d}_2 \tilde{\lambda}_2^{L_x-\ell + 1}
\left( 
\begin{array}{c} 
\lambda_2 \tilde{\bm{\chi}}_2\\
\tilde{\bm{\chi}}_2 \\
\end{array} 
\right),\nonumber \\
\end{eqnarray}
with the coefficients, $\tilde{d}_1$ and $\tilde{d}_2$, where the eigenvalues $\tilde{\lambda}_j$ satisfy $|\tilde{\lambda}_j|<1$, $j=1,2$. 
This solution is set to satisfy the Dirichlet boundary condition at the right edge, 
i.e., $\bm{\varphi}_{L_x+1} = 0$.
Then, following the same procedure as that for the left edge mode,
we can find the exact solution of the right edge mode by
setting $E$ so that the simultaneous eigenvector of $(A^{\dagger})^{-1}A$ and $(A^{\dagger})^{-1} (EI_2-B)$ exists,
and that the resulting $\tilde{\lambda}_{1}$ and $\tilde{\lambda}_{2}$ satisfy $|\tilde{\lambda}_{1}|, |\tilde{\lambda}_{2}|<1$.
We note that $(A^{\dagger})^{-1}A = (A^{-1}A^{\dagger})^{-1}$ holds, 
so these two matrices have common eigenvectors, $\bm{u}_j$,
and the corresponding eigenvalues are $\frac{1}{\mu_{j}}$.

\section{Exact solutions of the edge modes  \label{sec:disp_exact}}
In this appendix, we describe the derivation of the 
exact solutions of the edge modes 
for the QWZ-type model and the Haldane-type model.

\subsection{Qi-Wu-Zhang-type model  \label{sec:disp_exact_qwz}}
\subsubsection{Left edge mode}
In this model, the eigenvalues of $A^{-1} A^\dagger$ are given in Eq.~(\ref{eq:mu_qwz}). 
The corresponding eigenvectors are
\begin{eqnarray}
\bm{u}_{\pm}
= \begin{pmatrix}
i\alpha_1(t_1-\alpha_3) \\
\pm |\alpha_1| \sqrt{\alpha_3^2 -t_1^2} 
\end{pmatrix}. \label{eq:eigen_qwz}
\end{eqnarray}
Using (\ref{eq:eigen_qwz}), we can solve Eq.~(\ref{eq:E_to_solve}), and we have

\begin{eqnarray}
E^{\rm L}_{\pm} &=& 
-2m_0t_1 -2t_1 \cos k_y +2 t_2 \sin k_y \nonumber \\
&\pm& 2 \mathrm{sgn}(\alpha_1) \frac{\alpha_2 \sqrt{\alpha_3^2 -t_1^2} \sin k_y}{\alpha_3}.
 \nonumber \\ \label{eq:GQWZ_Left_edge}
\end{eqnarray}
From Eq.~(\ref{eq:GQWZ_Left_edge}), it is clear that 
the eigenenergy becomes complex-valued when $|\alpha_3| < |t_1|$,
i.e., the solution does not exist in this region,
which coincides with the discussion in the main text. 

To further determine the condition for the existence of the edge solution, 
we calculate the eigenvalues of $A^{-1}\left[EI -B \right]$.
After some algebras, we have
\begin{eqnarray}
\eta_\pm = \frac{P_{\pm} \mp Q_{\pm}}{\alpha_1(\alpha_3-t_1) \left(\alpha_1^2-\alpha_3^2+t_1^2 \right)} \label{eq:qwz_eta}
\end{eqnarray}
with 
\begin{widetext}
\begin{subequations}
\begin{eqnarray}
P_{\pm} = \alpha_1 \left(\alpha_3 - 
   t_1 \right) \left[(\alpha_3 - t_1) (-E^{\rm L}_{\pm} + 2 \alpha_3 m_0 + 2 \alpha_3  \cos k_y) + 
   2 (\alpha_1 \alpha_2 +\alpha_3t_2 - t_1t_2) \sin k_y \right], 
\end{eqnarray}
and
\begin{eqnarray}
Q_{\pm}=\sqrt{\alpha_3^2 - t_1^2}
  |\alpha_1| \left[\alpha_1 (E^{\rm L}_{\pm} + 2 \alpha_3 m_0 + 2 \alpha_3  \cos k_y) 
  - 2 (\alpha_2 \alpha_3 - t_1\alpha_2 + \alpha_1 t_2) \sin k_y \right].
\end{eqnarray}
\end{subequations}
\end{widetext}
Using (\ref{eq:mu_qwz}) and (\ref{eq:qwz_eta}),
we can obtain 
$\lambda^{\xi}_1$ and $\lambda_2^{\xi}$ for $\xi = \pm$,
and the edge solutions are obtained if $| \lambda^{\xi}_1|<1$ and $| \lambda^{\xi}_2|<1$
are satisfied for either $\xi = +$ or $\xi = -$. 

\subsubsection{Right edge mode}
The right edge modes can be obtained in the same manner as the left edge mode.
Here we summarize the results.
Firstly, the dispersion relation is given as
\begin{eqnarray}
E^{\rm R}_{\pm} = E^{\rm L}_{\pm}. 
\label{eq:GQWZ_Right_edge}
\end{eqnarray}
Secondly,
the eigenvalues of $(A^\dagger)^{-1}\left[EI -B \right]$ are
\begin{eqnarray}
\tilde{\eta}_\pm = \frac{\tilde{P}_{\pm} \pm \tilde{Q}_{\pm}}{\alpha_1(\alpha_3-t_1) \left(\alpha_1^2-\alpha_3^2+t_1^2 \right)} \label{eq:qwz_eta_tilde}
\end{eqnarray}
with 
\begin{widetext}
\begin{subequations}
\begin{eqnarray}
\tilde{P}_{\pm} = \alpha_1 \left(\alpha_3 - 
   t_1 \right) \left[(\alpha_3 - t_1) (-E^{\rm R}_{\pm} + 2 \alpha_3 m_0 + 2 \alpha_3  \cos k_y) -
   2 \left(\alpha_1 \alpha_2  -\alpha_3 t_2 + t_1t_2 \right)\sin k_y \right], 
\end{eqnarray}
and
\begin{eqnarray}
\tilde{Q}_{\pm}=\sqrt{\alpha_3^2 - t_1^2}
  |\alpha_1| \left[\alpha_1 (E^{\rm R}_{\pm} + 2 \alpha_3 m_0 + 2 \alpha_3 \cos k_y) +
   2 (\alpha_2 \alpha_3 - \alpha_2 t_1 -\alpha_1 t_2) \sin k_y \right].
\end{eqnarray}
\end{subequations}
\end{widetext}
Then, the edge solution can be found if 
$\tilde{\lambda}^{\xi}_1$ and $\tilde{\lambda}_2^{\xi}$ satisfy 
$| \tilde{\lambda}^{\xi}_1|<1$ and $|\tilde{\lambda}^{\xi}_2|<1$ for either $\xi = +$ or $\xi = -$. 

\subsection{Haldane-type model  \label{sec:disp_exact_hal}}
\subsubsection{Left edge mode}
For this model, the explicit forms of $A$ and $B$ are
\begin{eqnarray}
A= \left(
\begin{array}{cc}
X & 0 \\
t_1 & Y\\
\end{array}
\right),
\end{eqnarray}
and 
\begin{eqnarray}
B = \left(
\begin{array}{cc}
Z & V \\
V^{\ast}& W\\
\end{array}
\right),
\end{eqnarray}
where
\begin{subequations}
\begin{eqnarray}
X= t_2 \left[e^{-i \phi} + e^{i (\phi- k_y)} \right],
\end{eqnarray}
\begin{eqnarray}
Y= t_3 \left[e^{-i \phi^\prime} + e^{i (\phi^\prime- k_y)} \right],
\end{eqnarray}
\begin{eqnarray}
Z= 2 t_2 \cos \left( \phi + k_y \right),
\end{eqnarray}
\begin{eqnarray}
W = 2 t_3 \cos \left( \phi^\prime + k_y \right),
\end{eqnarray}
and 
\begin{eqnarray}
V = t_1 (1 + e^{-ik_y}).
\end{eqnarray}
\end{subequations}

For later use, we further introduce other four variables:
\begin{subequations}
\begin{eqnarray}
e^{i\theta_1}  = \frac{X^\ast}{X}, 
\end{eqnarray}
\begin{eqnarray}
e^{i\theta_2}  = \frac{Y^\ast}{Y}, 
\end{eqnarray}
\begin{eqnarray}
\alpha = \frac{t_1}{X}, 
\end{eqnarray}
and 
\begin{eqnarray}
\beta  = \frac{t_1}{Y}. 
\end{eqnarray}
\end{subequations}

The eigenvalues of $A^{-1}A^{\dagger}$ are 
\begin{eqnarray}
\mu_{\pm} &=& \frac{e^{i\theta_1} + e^{\theta_2} -\alpha \beta}{2}  \nonumber \\
&\pm &
\frac{\sqrt{\left(e^{i\theta_1} - e^{i\theta_2} \right)^2 + \alpha^2\beta^2 -2\alpha\beta\left(e^{i\theta_1} + e^{i\theta_2}\right)}}{2}, \nonumber \\
\end{eqnarray}
and the corresponding eigenvectors are 
\begin{eqnarray}
\bm{u}_{\pm} = \left( \begin{array}{c} \alpha\\ \mu_{\pm}-e^{i\theta_1}\\ \end{array} \right). \label{eq:eigen_upm}
\end{eqnarray}

The dispersion relation of the left edge mode can be determined by solving the linear equation of Eq.~(\ref{eq:E_to_solve}),
and the resulting dispersion relation is
\begin{eqnarray}
E^{\rm L}_{\pm} = \frac{G_\pm}{F_\pm} \label{eq:disp_left_gen}
\end{eqnarray}
with 
\begin{subequations}
\begin{eqnarray}
F_\pm = (X-Y) \alpha (\mu_\pm - e^{i\theta_1}) -t_1\alpha^2,
\end{eqnarray}
and 
\begin{eqnarray}
G_{\pm} =& -t_1\alpha^2 Z + (XW-YZ) \alpha(\mu_\pm - e^{i\theta_1}) \nonumber \\
+& \alpha^2 XV^\ast - \alpha (\mu_\pm - e^{i\theta_1}) t_1 V - YV(\mu_\pm - e^{i\theta_1})^2. 
\nonumber \\
\end{eqnarray}
\end{subequations}
Further, the eigenvalue of $A^{-1}(E_\pm I-B)$ can be obtained as
\begin{eqnarray}
\eta_{\pm} = \frac{[A^{-1}(E_\pm I-B)\bm{u}_{\pm}]_1}{[\bm{u}_{\pm}]_1}
=  \frac{E_\pm-Z}{X} - \frac{V(\mu_\pm-e^{i\theta_1})}{\alpha X}. \nonumber \\
\end{eqnarray}
We can then derive $\lambda^{\pm}_1$ and $\lambda_2^{\pm}$ by using Eqs.~(\ref{eq:lambda_1}) and (\ref{eq:lambda_2}).
If $| \lambda^{\xi}_1|<1$ and $| \lambda^{\xi}_2|<1$ 
are satisfied for either $\xi = +$ or $\xi = -$, 
we adopt $E_{\xi}$ as the dispersion of 
the left edge mode. 
\begin{figure}[t]
\begin{center}
\includegraphics[clip,width = \linewidth]{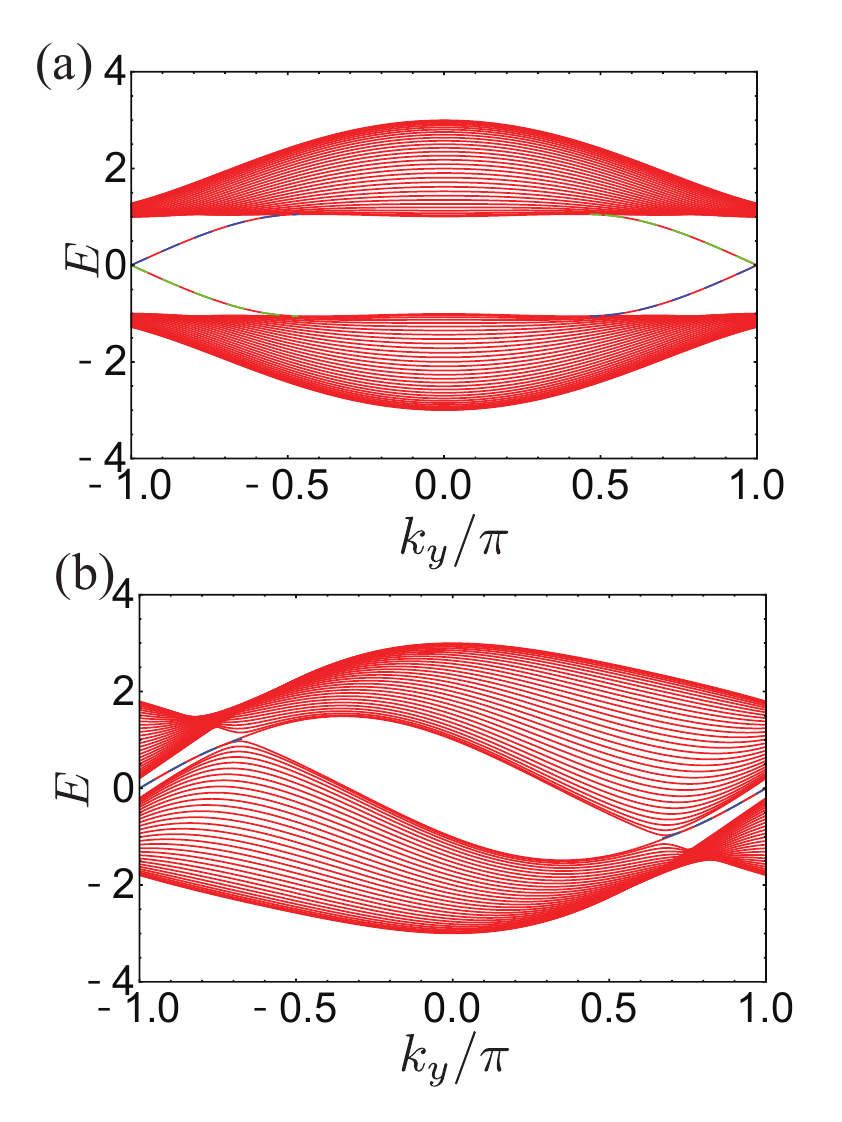}
\caption{
Dispersions relations for $(t_1 , t_2, t_3, \phi,\phi^\prime) = $ 
(a) $\left(1, 0.2, 0.2, \frac{\pi}{2}, -\frac{\pi}{2} \right)$ (i.e., the Haldane model with $\phi = \frac{\pi}{2}$) and
(b) $\left(1, 0.2, 0.2, \frac{\pi}{2}, \frac{\pi}{2} \right)$ (i.e., the Colom\'{e}s-Franz model model).
The colors of the lines indicate the same meaning as those in Fig.~\ref{fig:generic}.}
\label{fig:exact_HCF}
\end{center}
\vspace{-10pt}
\end{figure}

\subsubsection{Right edge mode \label{sec:right_edge}}
The right edge mode can be obtained in the same manner as the left edge mode, 
so we summarize only the results as follows: 
Firstly, the dispersion relation of the right edge mode 
is given as 
\begin{eqnarray}
E^{\rm R}_{\pm} = \frac{\tilde{G}_\pm}{\tilde{F}_\pm} \label{eq:disp_right_gen}
\end{eqnarray}
with 
\begin{subequations}
\begin{eqnarray}
\tilde{F}_{\pm} =  \alpha (X^\ast-Y^\ast) (\mu_\pm -e^{i\theta_1} ) + t_1(\mu_\pm -e^{i\theta_1})^2,\nonumber \\
\end{eqnarray}
and 
\begin{eqnarray}
\tilde{G}_\pm &=&\alpha^2 X^\ast V^\ast + \alpha(X^\ast W - Y^\ast Z +t_1 V^\ast) (\mu_\pm -e^{i\theta_1} ) \nonumber \\
&-&(Y^\ast V-t_1 W)(\mu_\pm -e^{i\theta_1} )^2. \nonumber \\
\end{eqnarray}
\end{subequations}
Secondly, the eigenvalues of $\left(A^\dagger \right)^{-1}(E_\pm I-B)$ are
\begin{eqnarray}
\tilde{\eta}_{\pm} &=& \frac{Y^\ast (E^{\rm R}_\pm -Z) + t_1 V^\ast}{X^\ast Y^\ast} \nonumber \\
&-&\frac{[Y^\ast V+t_1 (E^{\rm R}_\pm-W)](\mu_\pm -e^{i\theta_1})}{\alpha X^\ast Y^\ast }. \nonumber \\
\end{eqnarray}
The solution can be found if 
$\tilde{\lambda}^{\xi}_1$ and $\tilde{\lambda}_2^{\xi}$ satisfy $| \tilde{\lambda}^{\xi}_1|<1$ and $|\tilde{\lambda}^{\xi}_2|<1$ for either $\xi = +$ or $\xi = -$. 

\subsubsection{Dispersion relations for some model parameters \label{sec:disp_exact_2}}
Although the dispersion relations of 
Eqs.~(\ref{eq:disp_left_gen}) and (\ref{eq:disp_right_gen}) are exact for generic parameters, 
their forms are fairly complicated. 
Nevertheless, the simple forms of the dispersion relations can 
be obtained for some special model parameters.

Firstly, for the original Haldane model 
with $\phi = \pi/2$, $\phi^\prime = -\pi/2$ and $t_2 = t_3$,
 the dispersion relation of the left edge mode and of the right edge mode are given as 
\begin{eqnarray}
 E^{\rm L}  (k_y) =  - E^{\rm R}  (k_y) 
 = - \frac{6|t_1|t_2 \sin k_y}{\sqrt{t_1^2 + 8t_2^2 (1-\cos k_y)}}. \nonumber \\ \label{eq:disp_chiral}
\end{eqnarray}
Note that this result was already obtained under a certain \textit{Ansatz} in Ref.~\onlinecite{Huang2012} .

Secondly, for the Colom\'{e}s-Franz model with $\phi =\phi^\prime$ and $t_2 = t_3$, 
the dispersion relations of the left edge mode and of the right edge mode are
\begin{eqnarray}
 E^{\rm L}  (k_y) =  E^{\rm R}  (k_y) 
 = -2t_2 \cos \phi (2 + \cos k_y) - 6t_2 \sin \phi \sin k_y. \nonumber \\ \label{eq:disp_antichiral}
\end{eqnarray}
It is worth noting that $E  (k_y)$ does not depend on $t_1$, unlike the case of the Haldane model.

The comparison with the numerical results and the present exact solutions 
is shown in Fig.~\ref{fig:exact_HCF},
where we see good agreement.


\begin{thebibliography}{1}
\bibitem{Hasan2010}
M. Z. Hasan and C. L. Kane, Rev. Mod. Phys. \textbf{82}, 3045 (2010).

\bibitem{Qi2011}
X.-L. Qi and S.-C. Zhang, Rev. Mod. Phys. \textbf{83}, 1057 (2011).

\bibitem{Halperin1982}
B. I. Halperin, 
Phys. Rev. B \textbf{25}, 2185 (1982).

\bibitem{Hatsugai1993}
Y. Hatsugai,
Phys. Rev. Lett. \textbf{71}, 3697 (1993).

\bibitem{Mizoguchi2019}
T. Mizoguchi and T. Koma, 
Phys. Rev. B \textbf{99}, 184418 (2019).

\bibitem{Elbau2002}
P. Elbau and G. M. Graf,
Commun. Math. Phys. \textbf{229}, 415 (2002).

\bibitem{Hsieh2008}
D. Hsieh, D. Qian, L. Wray, Y. Xia, Y. S. Hor, R. J. Cava, and 
M. Z. Hasan, Nature \textbf{452}, 970 (2008).

\bibitem{Hsieh2009}
D. Hsieh, Y. Xia, L. Wray, D. Qian, A. Pal, J. H. Dil, J. Osterwalder, 
F. Meier, G. Bihlmayer, C. L. Kane, Y. S. Hor, R. J. Cava, and M. Z. Hasan, 
Science \textbf{323}, 919 (2009).

\bibitem{Mao2011}
S. Mao and Y. Kuramoto,
Phys. Rev. B \textbf{83}, 085114 (2011).

\bibitem{Xu2014}
Y. Xu, R.-L. Chu, and C. Zhang,
Phys. Rev. Lett. \textbf{112}, 136402 (2014).

\bibitem{Ying2018}
X. Ying and A. Kamenev,
Phys. Rev. Lett. \textbf{121}, 086810 (2018).

\bibitem{Bahari2019}
M. Bahari and M. V. Hosseini,
Phys. Rev. B \textbf{99}, 155128 (2019).

\bibitem{Upreti2019}
L. K. Upreti, C. Evain, S. Randoux, P. Suret, A. Amo, and P. Delplace,
arXiv:1907.09914.

\bibitem{Murakami2007}
S. Murakami, New J. Phys. \textbf{9}, 356 (2007).

\bibitem{Vafek2014}
O. Vafek and A. Vishwanath, 
Annu. Rev. Condens. Matter Phys. \textbf{5}, 83 (2014).

\bibitem{Armitage2018}
N. P. Armitage, E. J. Mele, and 
A. Vishwanath, Rev. Mod. Phys.
\textbf{90}, 015001 (2018).

\bibitem{Fang2015}
C. Fang, Y. Chen, H.-Y. Kee, and L. Fu, 
Phys. Rev. B \textbf{92}, 081201(R) (2015).

\bibitem{Yamakage2016}
A. Yamakage, Y. Yamakawa, Y. Tanaka, and Y. Okamoto,
J. Phys. Soc. Jpn. \textbf{85}, 013708 (2016).

\bibitem{Colomes2018} 
E. Colom\'{e}s and M. Franz,
Phys. Rev. Lett. \textbf{120}, 086603 (2018).

\bibitem{Vila2019}
M. Vila, N. T. Hung, S. Roche, and R. Saito,
Phys. Rev. B \textbf{99}, 161404(R) (2019).

\bibitem{Mandal2019} 
S. Mandal, R. Ge, and T. C. H.  Liew,
Phys. Rev. B \textbf{99}, 115423 (2019).

\bibitem{Wang2020}
C. Wang, L. Zhang, P. Zhang, J. Song, and Y.-X. Li,
Phys. Rev. B \textbf{101}, 045407 (2020).

\bibitem{Bhowmick2020}
D. Bhowmick and P. Sengupta,
Phys. Rev. B \textbf{101}, 195133 (2020).

\bibitem{Denner2020}
M. M. Denner, J. L. Lado, and O. Zilberberg,
Phys. Rev. Research \textbf{2}, 043190 (2020).

\bibitem{Yang2020}
Y. Yang, D. Zhu, Z. Hang, and Y. Chong, 
Sci. China Phys. Mech. Astron. \textbf{64}, 257011 (2021).

\bibitem{Chen2020}
J. Chen, W. Liang, and Z.-Y. Li,
Phys. Rev. B \textbf{101}, 214102 (2020).

\bibitem{Zhou2020}
P. Zhou, G.-G. Liu, Y. Yang, Y.-H. Hu, S. Ma, H. Xue, Q. Wang, L. Deng, and B. Zhang, 
Phys. Rev. Lett. \textbf{125}, 263603 (2020).

\bibitem{Prodan2016}
E. Prodan and H. Schulz-Baldes, 
\textit{Bulk and boundary invariants for complex topological insulators: 
From K-theory to physics} (Springer Int. Pub., Szwitzerland, 2016).

\bibitem{Graf2018}
G. M. Graf and J. Shapiro, 
Commun. Math. Phys. \textbf{363}, 829 (2018).

\bibitem{Lee1981}
D. H. Lee and J. D. Joannopoulos,
Phys. Rev. B \textbf{23}, 4988 (1981). 

\bibitem{Hatsugai1993_2}
Y. Hatsugai, 
Phys. Rev. B \textbf{48}, 11851 (1993). 

\bibitem{Molinari1997}
L. Molinari,
J. Phys. A: Math. Gen. \textbf{30}, 983 (1997).

\bibitem{Schulz-Baldes2000}
H. Schulz-Baldes, J. Kellendonk, and T. Richter,
J. Phys. \textbf{A33} L27-L32 (2000). 

\bibitem{Kellendonk2002}
J. Kellendonk, T. Richter, and H. Schulz-Baldes,
Rev. Math. Phys. \textbf{14}, 87-119 (2002).

\bibitem{Teo2008}
J. C. Y. Teo, L. Fu, and C. L. Kane,
Phys. Rev. B \textbf{78}, 045426 (2008).

\bibitem{Mao2010}
S. Mao, Y. Kuramoto, K.-I. Imura, and A. Yamakage,
J. Phys. Soc. Jpn. \textbf{79}, 124709 (2010).

\bibitem{Huang2012}
Z. Huang and D. P. Arovas, 
arXiv:1205.6266.

\bibitem{Doh2014}
H. Doh, G. S. Jeon, and H. J. Choi,
arXiv:1408.4507.

\bibitem{Dwivedi2016}
V. Dwivedi and V. Chua,
Phys. Rev. B \textbf{93}, 134304 (2016).

\bibitem{Pantaleon2017}
P. A. Pantale\'{o}n and Y. Xian,
J. Phys.: Condens. Matter \textbf{29}, 295701 (2017).

\bibitem{Kunst2019}
F. K. Kunst, G. van Miert, and E. J. Bergholtz, 
Phys. Rev. B \textbf{99}, 085426 (2019); 
Phys. Rev. B \textbf{99}, 085427 (2019).

\bibitem{Mizoguchi2020} 
T. Mizoguchi, T. Koma, and Y. Yoshida,
Phys. Rev. B \textbf{101}, 014442 (2020). 

\bibitem{Qi2006}
X.-L. Qi, Y.-S. Wu, and S.-C. Zhang,
Phys. Rev. B \textbf{74}, 085308 (2006).

\bibitem{Haldane1988}
F. D. M. Haldane, 
Phys. Rev. Lett. \textbf{61}, 2015 (1988). 

\bibitem{Yakovenko1990}
V. M. Yakovenko, 
Phys. Rev. Lett. \textbf{65}, 251 (1990).

\bibitem{Fukui2005}
T. Fukui, Y. Hatsugai, and H. Suzuki,
J. Phys. Soc. Jpn. 
\textbf{74}, 1674 (2005).

\end{thebibliography}
\end{document}